\documentclass[pra,twocolumn,showpacs,preprintnumbers,superscriptaddress]
{revtex4}

\usepackage{times}
\usepackage{bm}
\usepackage{graphicx}
\usepackage{amsbsy}
\usepackage{amsmath}
\usepackage{amsfonts}
\usepackage{amsthm}

\begin{document}

\theoremstyle{plain}
\newtheorem{theorem}{Theorem}
\newtheorem{lemma}[theorem]{Lemma}
\newtheorem{corollary}[theorem]{Corollary}
\newtheorem{proposition}[theorem]{Proposition}
\newtheorem{conjecture}[theorem]{Conjecture}

\theoremstyle{definition}
\newtheorem{definition}[theorem]{Definition}

\theoremstyle{remark}
\newtheorem*{remark}{Remark}
\newtheorem{example}{Example}

\title{An analytical condition for the violation of Mermin's inequality by any three qubit state}

\author{Satyabrata Adhikari}
\thanks{tapisatya@gmail.com}
\affiliation{Birla Institute of Technology Mesra, Ranchi-835215,
India}

\author{A. S. Majumdar}
\thanks{archan@bose.res.in}
\affiliation{S. N. Bose National Centre for Basic Sciences, Block
JD, Sector-III, Salt Lake, Kolkata-700098, India}
\date{\today}

\begin{abstract}
Mermin's inequality is the generalization of the Bell-CHSH
inequality for three qubit states. The violation of the Mermin
inequality guarantees the fact that there exists quantum
non-locality either between two or three qubits in a three
qubit system. In the absence of an analytical result to this effect, in
order to check for the violation of Mermin's inequality one has to perform a
numerical optimization procedure for even three qubit pure states.  Here we derive an
analytical formula for the maximum value of
the expectation of the Mermin operator in terms of eigenvalues of
symmetric matrices, that gives the maximal violation of the
Mermin inequality for all three qubit pure and mixed
states.
\end{abstract}

 \maketitle

\section{Introduction}

The impossiblity of reproducing the effect of quantum correlations between the
outcomes of the distant measurements using local hidden variable
theories is known as quantum non-locality. In
1964, Bell constructed an inequality which is satisfied in the absence of
non-local correlations between the results of distant measurements \cite{bell}.
Experimental
violation of  Bell's inequality confirm the existence of the
non-local correlation between the outcome of the measurements.
 The most well-known
form of the Bell inequality is given by Clauser, Horne, Shimony, and
Holt (CHSH) and it is known as Bell-CHSH inequality \cite{chsh}.
The Bell-CHSH operator for two qubits  is given by
 $B_{CHSH} = \hat{a}.\vec{\sigma}\otimes
(\hat{b}+\hat{b'}).\vec{\sigma}+\hat{a'}.\vec{\sigma}\otimes
(\hat{b}-\hat{b'}).\vec{\sigma}$,
 where $\hat{a}$, $\hat{a'}$, $\hat{b}$, $\hat{b'}$ are unit
vectors in $R^{3}$.
The Bell-CHSH inequality is then given by
$|\langle B_{CHSH}\rangle_{\rho}|\leq 2$,
where $\rho$ denotes any two qubit pure or mixed state.
This inequality is violated by any
two qubit pure entangled state, but on the contrary not all two
qubit
mixed entangled states violate the  Bell-CHSH inequality.

Foundational interest in nonlocality is bolstered through its connection with
information theoretic tasks such as teleportation \cite{tele}. Quantum nonlocality
finds applications
in several information theoretic protocols such as device independent quantum key
generation \cite{diqkd}, quantum state estimation \cite{stateestim}, and communication
complexity \cite{commcomp}, where the amount of violation of the  Bell-CHSH inequality
is important.
In order to
obtain the maximal violation of the Bell-CHSH inequality,
one has to calculate the expectation of the
Bell-CHSH operator by maximizing over all measurements of spin in
the directions $\hat{a}$, $\hat{a'}$, $\hat{b}$, $\hat{b'}$. Therefore, the
problem of maximal violation of the Bell-CHSH inequality reduces
to an optimization problem. The optimization problem for the two
qubit system was analytically solved by Horodecki \cite {horo}
by expressing the value of $\langle B_{max}\rangle_{\rho}$ in
terms of the eigenvalues of the symmetric matrix
$T_{\rho}^{T}T_{\rho}$, where $T_{\rho}$ is the correlation matrix
of the state $\rho$. Therefore, the maximal violation of the
Bell-CHSH inequality depends on the eigenvalues of the symmetric
matrix
$T_{\rho}^{T}T_{\rho}$.

Like two qubit non-locality, non-locality for three qubit systems has also
been studied using various approaches.
A generalized form of the Bell-CHSH inequality was obtained for three
qubits called Mermin's inequality \cite{mermin} which can be violated by not
only genuine entangled three qubit states but also by biseparable
states. On the other hand, all genuine entangled three qubit states
violate the Svetlichny inequality \cite{svet}. There has been quite a bit
of recent interest in studying the nonlocality of tripartite systems. A notable
direction in this context is the so-called `superactivation of nonlocality' \cite{super}
which has been investigated also for the case of three qubits \cite{super2,sarkar}. The
relation
of nonlocality with quantum uncertainty has been exhibited for tripartite systems
using fine-graining \cite{fur1}, in the context of biased \cite{fur2} and unbiased
qauntum games. The security of quantum cryptography is connected with quantum
nonlocality \cite{gisin}, that is especially relevant in the context of device
independent quantum key distribution.

Violation of the Mermin inequality has been computed for several
three qubit states such as GHZ and W-states earlier \cite{scarani,
scarani1, emary, chi}. In order to find the maximum violation of
the Mermin inequality for three qubit states one has to tackle the
optimization problem numerically because there does not exist any
analytical formula  for even pure three qubit states. Motivated by
the work of Horodecki \cite{horo} in the context of two qubit
systems, in the present work we perform the optimization problem
involved in the Mermin inequality analytically and obtain a
formula for the maximal value of the expectation of the Mermin
operator in terms of the eigenvalues of  symmetric matrices, that
gives the maximal violation of the Mermin inequality not only for
pure states but also for mixed states. The plan of this paper is
as follows. In section-II, we solve the optimization problem
analytically and obtain the maximum value of the expectation of
the Mermin operator in terms of eigenvalues. In section-III,  we
provide a few examples where the magnitude of the Mermin operator
is calculated using our derived formular for pure and mixed
states. Certain concluding remarks are presented in section-IV.

\section{Maximum value of the expectation of the Mermin operator in terms of eigenvalues}

Let $B_{M}$ be the Mermin operator defined as \cite{chi}
\begin{eqnarray}
B_{M}&=&\hat{a}_{1}.\vec{\sigma}\otimes
\hat{a}_{2}.\vec{\sigma}\otimes
\hat{a}_{3}.\vec{\sigma}-\hat{a}_{1}.\vec{\sigma}\otimes
\hat{b}_{2}.\vec{\sigma}\otimes \hat{b}_{3}.\vec{\sigma}
\nonumber\\&-&\hat{b}_{1}.\vec{\sigma}\otimes
\hat{a}_{2}.\vec{\sigma}\otimes
\hat{b}_{3}.\vec{\sigma}-\hat{b}_{1}.\vec{\sigma}\otimes
\hat{b}_{2}.\vec{\sigma}\otimes
\hat{a}_{3}.\vec{\sigma}\label{merminoperator}
\end{eqnarray}
where $\hat{a}_{j}$ and $\hat{b}_{j}$ (j=1,2,3) are unit vectors
in $R^{3}$, and $\vec{\sigma}=(\sigma_{x},\sigma_{y},\sigma_{z})$
is the vector of the Pauli matrices.\\
For any three qubit state $\rho$, the Mermin inequality is
\begin{eqnarray}
\langle B_{M}\rangle_{\rho}\leq 2\label{mermininequality}
\end{eqnarray}
where
\begin{eqnarray}
\rho&=&\frac{1}{8}[I\otimes I\otimes I+\vec{l}.\vec{\sigma}\otimes
I\otimes I+I\otimes \vec{m}.\vec{\sigma}\otimes
I\nonumber\\&+&I\otimes I\otimes
\vec{n}.\vec{\sigma}+\vec{u}.\vec{\sigma}\otimes
\vec{v}.\vec{\sigma}\otimes I+\vec{u}.\vec{\sigma}\otimes
I\otimes\vec{w}.\vec{\sigma}\nonumber\\&+&I\otimes
\vec{v}.\vec{\sigma}\otimes
\vec{w}.\vec{\sigma}+\sum_{i,j,k=x,y,z}t_{ijk}\sigma_{i}\otimes
\sigma_{j}\otimes \sigma_{k} ]\label{threequbitstate}
\end{eqnarray}
with
\begin{eqnarray}
&&l_{i}=Tr(\rho(\sigma_{i}\otimes I\otimes I)),
m_{i}=Tr(\rho(I\otimes \sigma_{i}\otimes I)),\nonumber\\&&
n_{i}=Tr(\rho(I \otimes I\otimes \sigma_{i})),(i=x,y,z)
\label{coefficient1}
\end{eqnarray}
\begin{eqnarray}
&& u_{i}v_{i}=Tr(\rho(\sigma_{i}\otimes \sigma_{i} \otimes
I)),u_{i}w_{i}=Tr(\rho(\sigma_{i}\otimes I \otimes
\sigma_{i})),\nonumber\\&& v_{i}w_{i}=Tr(\rho(I \otimes \sigma_{i}
\otimes \sigma_{i})), (i=x,y,z)
\label{coefficient2}
\end{eqnarray}
\begin{eqnarray}
t_{ijk}=Tr(\rho(\sigma_{i}\otimes \sigma_{j}\otimes \sigma_{k})),
(i,j,k=x,y,z) \label{coefficient3}
\end{eqnarray}
We will now derive the necessary and sufficient condition which
tells us that when a three-qubit state $\rho$ violates Mermin's inequality.
The expectation value of the Mermin operator with respect to the
state $\rho$ given by (\ref{threequbitstate}) is
\begin{eqnarray}
\langle
B_{M}\rangle_{\rho}&=&Tr(B_{M}\rho)\nonumber\\&=&\sum_{i,j,k=x,y,z}a_{1i}a_{2j}a_{3k}t_{ijk}\nonumber\\&-&
\sum_{i,j,k=x,y,z}a_{1i}b_{2j}b_{3k}t_{ijk}\nonumber\\&-&
\sum_{i,j,k=x,y,z}b_{1i}a_{2j}b_{3k}t_{ijk}\nonumber\\&-&
\sum_{i,j,k=x,y,z}b_{1i}b_{2j}a_{3k}t_{ijk}
\nonumber\\&=&(\hat{a}_{1},\hat{a}_{3}^{T}\vec{T}\hat{a}_{2})
-(\hat{a}_{1},\hat{b}_{3}^{T}\vec{T}\hat{b}_{2})\nonumber\\&-&(\hat{b}_{1},\hat{b}_{3}^{T}\vec{T}\hat{a}_{2})
-(\hat{b}_{1},\hat{a}_{3}^{T}\vec{T}\hat{b}_{2})\label{expectationvalue}
\end{eqnarray}
where $\hat{a}_{s}=(a_{sx},a_{sy},a_{sz})$ and
$\hat{b}_{s}=(b_{sx},b_{sy},b_{sz})$, (s=1,2,3) and
$(\vec{x},\vec{y})$ denotes the inner product of two vectors
$\vec{x}$ and $\vec{y}$ and is defined as
$(\vec{x},\vec{y})=\|x\|\|y\|cos\theta$, $\theta$ being the angle
between $\vec{x}$ and $\vec{y}$. The superscript $T$ refers to
transpose, and $\vec{T}=(T_{x},T_{y},T_{z})$;
$T_{x}=\begin{pmatrix}
  t_{xxx} & t_{xyx} & t_{xzx} \\
  t_{xxy} & t_{xyy} & t_{xzy} \\
  t_{xxz} & t_{xyz} & t_{xzz}
\end{pmatrix}$, $T_{y}=\begin{pmatrix}
  t_{yxx} & t_{yyx} & t_{yzx} \\
  t_{yxy} & t_{yyy} & t_{yzy} \\
  t_{yxz} & t_{yyz} & t_{yzz}
\end{pmatrix}$,$T_{z}=\begin{pmatrix}
  t_{zxx} & t_{zyx} & t_{zzx} \\
  t_{zxy} & t_{zyy} & t_{zzy} \\
  t_{zxz} & t_{zyz} & t_{zzz}
\end{pmatrix}$\\\\
\textbf{Theorem-1:} If the symmetric matrices
$T_{x}^{T}T_{x}$, $T_{y}^{T}T_{y}$ and $T_{z}^{T}T_{z}$ have unique
largest eigenvalues $\lambda_{x}^{max}$,$\lambda_{y}^{max}$ and
$\lambda_{z}^{max}$ respectively, then Mermin's inequality is
violated if
\begin{eqnarray}
\langle B_{M}^{max}\rangle_{\rho}&=&max_{B_{M}}Tr(B_{M}\rho)
\nonumber\\&=&max\{2\sqrt{\lambda_{x}^{max}},2\sqrt{\lambda_{y}^{max}}
,2\sqrt{\lambda_{z}^{max}}\}\nonumber\\&>&2\label{theorem1}
\end{eqnarray}
\textbf{Proof:} In the expression for $\langle
B_{M}\rangle_{\rho}$ given by Eq.(\ref{expectationvalue}), we
first
 simplify the vectors
$\hat{a}_{3}^{T}\vec{T}\hat{a}_{2}$,
$\hat{b}_{3}^{T}\vec{T}\hat{b}_{2}$,
$\hat{b}_{3}^{T}\vec{T}\hat{a}_{2}$,
$\hat{a}_{3}^{T}\vec{T}\hat{b}_{2}$. We choose the vectors
$\hat{a}_{2}$, $\hat{a}_{3}$, $\hat{b}_{2}$ and $\hat{b}_{3}$ in
such a way that they maximize the quantity $\langle
B_{M}\rangle_{\rho}$ over all the operators $B_{M}$. Let us proceed  by considering the
following cases sequentially:\\
\textbf{Case-I:} In this case we choose the vectors in such a way that the maximized
expectation value of the Mermin operator is given by
 $\langle B_{M}^{(1)}\rangle_{\rho}=2\sqrt{\lambda_{x}^{max}}$.\\
(i) The vector $\hat{a}_{3}^{T}\vec{T}\hat{a}_{2}$ can be
simplified as
\begin{eqnarray}
\hat{a}_{3}^{T}\vec{T}\hat{a}_{2}&=&((\hat{a}_{3},T_{x}\hat{a}_{2}),(\hat{a}_{3},T_{y}\hat{a}_{2}),(\hat{a}_{3},T_{z}\hat{a}_{2}))
\nonumber\\&=&(\|\hat{a}_{3}^{max}\|\|T_{x}\hat{a}_{2}\|,0,0)
\label{vector1}
\end{eqnarray}
where $\hat{a}_{3}^{max}$ is a unit vector along
$T_{x}\hat{a}_{2}$ and perpendicular to $T_{y}\hat{a}_{2}$ and
$T_{z}\hat{a}_{2}$.
Since $\hat{a}_{3}^{max}$ is a unit vector,
$\|\hat{a}_{3}^{max}\|=1$. Again,
$\|T_{x}\hat{a}_{2}\|^{2}=(T_{x}\hat{a}_{2},T_{x}\hat{a}_{2})=(\hat{a}_{2},T_{x}^{T}T_{x}\hat{a}_{2})$.
If $\lambda_{x}^{max}$ is the largest eigenvalue of the symmetric
matrix $T_{x}^{T}T_{x}$ and $\hat{a}_{2}^{max}$ is the
corresponding unit vector, then
$\|T_{x}\hat{a}_{2}\|^{2}=(\hat{a}_{2},T_{x}^{T}T_{x}\hat{a}_{2})=(\hat{a}_{2},\lambda_{x}^{max}\hat{a}_{2}^{max})$.
If $\hat{a}_{2}$ is the unit vector along $\hat{a}_{2}^{max}$, then
$\|T_{x}\hat{a}_{2}\|^{2}=\lambda_{x}^{max}$. Thus,
\begin{eqnarray}
\hat{a}_{3}^{T}\vec{T}\hat{a}_{2}=(\sqrt{\lambda_{x}^{max}},0,0)
\label{vector1a}
\end{eqnarray}
(ii) The vector $\hat{a}_{3}^{T}\vec{T}\hat{b}_{2}$ can be
simplified as
\begin{eqnarray}
\hat{a}_{3}^{T}\vec{T}\hat{b}_{2}&=&((\hat{a}_{3},T_{x}\hat{b}_{2}),(\hat{a}_{3},T_{y}\hat{b}_{2}),(\hat{a}_{3},T_{z}\hat{b}_{2}))
\nonumber\\&=&(0,-\|\hat{a}_{3}^{min}\|\|T_{y}\hat{b}_{2}\|,0)
\label{vector2}
\end{eqnarray}
where $\hat{a}_{3}^{min}$ is the unit vector antiparallel to
$T_{y}\hat{b}_{2}$ and perpendicular to $T_{x}\hat{b}_{2}$ and
$T_{z}\hat{b}_{2}$.
Since $\hat{a}_{3}^{min}$ is a unit vector,
$\|\hat{a}_{3}^{min}\|=1$. Again,
$\|T_{y}\hat{b}_{2}\|^{2}=(T_{y}\hat{b}_{2},T_{y}\hat{b}_{2})=(\hat{b}_{2},T_{y}^{T}T_{y}\hat{b}_{2})$.
If $\lambda_{y}^{max}$ is the largest eigenvalue of the symmetric
matrix $T_{y}^{T}T_{y}$ and $\hat{b}_{2}^{max}$ is the
corresponding unit vector, then
$\|T_{y}\hat{b}_{2}\|^{2}=(\hat{b}_{2},T_{y}^{T}T_{y}\hat{b}_{2})=(\hat{b}_{2},\lambda_{y}^{max}\hat{b}_{2}^{max})$.
If $\hat{b}_{2}$ is the unit vector along $\hat{b}_{2}^{max}$, then
$\|T_{y}\hat{b}_{2}\|^{2}=\lambda_{y}^{max}$. Thus,
\begin{eqnarray}
\hat{a}_{3}^{T}\vec{T}\hat{b}_{2}=(0,-\sqrt{\lambda_{y}^{max}},0)
\label{vector2a}
\end{eqnarray}
(iii) The vector $\hat{b}_{3}^{T}\vec{T}\hat{b}_{2}$ can be
simplified as
\begin{eqnarray}
\hat{b}_{3}^{T}\vec{T}\hat{b}_{2}&=&((\hat{b}_{3},T_{x}\hat{b}_{2}),(\hat{b}_{3},T_{y}\hat{b}_{2}),(\hat{b}_{3},T_{z}\hat{b}_{2}))
\nonumber\\&=&(0,-\|\hat{b}_{3}^{min}\|\|T_{y}\hat{b}_{2}\|,0)\nonumber\\&=&(0,-\sqrt{\lambda_{y}^{max}},0)
\label{vector3}
\end{eqnarray}
where $\hat{b}_{3}^{min}$ is the unit vector antiparallel to
$T_{y}\hat{b}_{2}$ and perpendicular to $T_{x}\hat{b}_{2}$ and
$T_{z}\hat{b}_{2}$. Since $\hat{b}_{3}^{min}$ is a unit vector,
$\|\hat{b}_{3}^{min}\|=1$.\\
(iv) The vector $\hat{b}_{3}^{T}\vec{T}\hat{a}_{2}$ can be
simplified as
\begin{eqnarray}
\hat{b}_{3}^{T}\vec{T}\hat{a}_{2}&=&((\hat{b}_{3},T_{x}\hat{a}_{2}),(\hat{b}_{3},T_{y}\hat{a}_{2}),(\hat{b}_{3},T_{z}\hat{a}_{2}))
\nonumber\\&=&(\|\hat{b}_{3}^{max}\|\|T_{x}\hat{a}_{2}\|,0,0)=(\sqrt{\lambda_{x}^{max}},0,0)
\label{vector4}
\end{eqnarray}
where $\hat{b}_{3}^{max}$ is the unit vector along
$T_{x}\hat{a}_{2}$ and perpendicular to $T_{y}\hat{a}_{2}$ and
$T_{z}\hat{a}_{2}$.\\
From
(\ref{expectationvalue}),(\ref{vector1a}),(\ref{vector2a}),(\ref{vector3}),(\ref{vector4}),
we have
\begin{eqnarray}
\langle
B_{M}^{(1)}\rangle_{\rho}&=&max_{\hat{a}_{1},\hat{b}_{1}}[(\hat{a}_{1},(\sqrt{\lambda_{x}^{max}},0,0))
\nonumber\\&+&
(\hat{a}_{1},(0,\sqrt{\lambda_{y}^{max}},0))-(\hat{b}_{1},(\sqrt{\lambda_{x}^{max}},0,0))
\nonumber\\&+&(\hat{b}_{1},(0,\sqrt{\lambda_{y}^{max}},0))]\nonumber\\&=&\|\hat{a}_{1}^{max}\|\|(\sqrt{\lambda_{x}^{max}},0,0)\|
\nonumber\\&+&\|\hat{b}_{1}^{max}\|\|(\sqrt{\lambda_{x}^{max}},0,0)\|\nonumber\\&=&2\sqrt{\lambda_{x}^{max}}
\label{result1}
\end{eqnarray}
$\hat{a}_{1}^{max}$ is the unit vector parallel to
$(\sqrt{\lambda_{x}^{max}},0,0)$ and perpendicular to
$(0,\sqrt{\lambda_{y}^{max}},0)$; $\hat{b}_{1}^{max}$ is the
unit vectors antiparallel to $(\sqrt{\lambda_{x}^{max}},0,0)$ and perpendicular to
$(0,\sqrt{\lambda_{y}^{max}},0)$.\\

\textbf{Case-II:} In this case we choose the vectors in such a way
that the maximized expectation value of the Mermin operator is given by
 $\langle B_{M}^{(2)}\rangle_{\rho}=2\sqrt{\lambda_{y}^{max}}$.\\
(i) The vector $\hat{a}_{3}^{T}\vec{T}\hat{a}_{2}$ can be
simplified as
\begin{eqnarray}
\hat{a}_{3}^{T}\vec{T}\hat{a}_{2}&=&((\hat{a}_{3},T_{x}\hat{a}_{2}),(\hat{a}_{3},T_{y}\hat{a}_{2}),(\hat{a}_{3},T_{z}\hat{a}_{2}))
\nonumber\\&=&(0,\|\hat{a}_{3}^{max}\|\|T_{y}\hat{a}_{2}\|,0)
\label{vector11}
\end{eqnarray}
where $\hat{a}_{3}^{max}$ is the unit vector along
$T_{y}\hat{a}_{2}$ and perpendicular to $T_{x}\hat{a}_{2}$ and
$T_{z}\hat{a}_{2}$.
Repeating the steps of Case-I, we find
\begin{eqnarray}
\hat{a}_{3}^{T}\vec{T}\hat{a}_{2}=(0,\sqrt{\lambda_{y}^{max}},0)
\label{vector11a}
\end{eqnarray}
(ii) Similarly, we obtain
\begin{eqnarray}
\hat{a}_{3}^{T}\vec{T}\hat{b}_{2}=(-\sqrt{\lambda_{x}^{max}},0,0)
\label{vector22a}
\end{eqnarray},
(iii) and
\begin{eqnarray}
\hat{b}_{3}^{T}\vec{T}\hat{b}_{2}&=&((\hat{b}_{3},T_{x}\hat{b}_{2}),(\hat{b}_{3},T_{y}\hat{b}_{2}),(\hat{b}_{3},T_{z}\hat{b}_{2}))
\nonumber\\&=&(-\|\hat{b}_{3}^{min}\|\|T_{x}\hat{b}_{2}\|,0,0)\nonumber\\&=&(-\sqrt{\lambda_{x}^{max}},0,0)
\label{vector33}
\end{eqnarray},
(iv) and
\begin{eqnarray}
\hat{b}_{3}^{T}\vec{T}\hat{a}_{2}&=&((\hat{b}_{3},T_{x}\hat{a}_{2}),(\hat{b}_{3},T_{y}\hat{a}_{2}),(\hat{b}_{3},T_{z}\hat{a}_{2}))
\nonumber\\&=&(0,\|\hat{b}_{3}^{max}\|\|T_{y}\hat{a}_{2}\|,0)=(0,\sqrt{\lambda_{y}^{max}},0)
\label{vector44}
\end{eqnarray}
where $\hat{b}_{3}^{max}$ is the unit vector along
$T_{y}\hat{a}_{2}$ and perpendicular to $T_{x}\hat{a}_{2}$ and
$T_{z}\hat{a}_{2}$.

Now, from
(\ref{expectationvalue}),(\ref{vector11a}),(\ref{vector22a}),(\ref{vector33}),(\ref{vector44}),
we have
\begin{eqnarray}
\langle
B_{M}^{(2)}\rangle_{\rho}&=&max_{\hat{a}_{1},\hat{b}_{1}}[(\hat{a}_{1},(0,\sqrt{\lambda_{y}^{max}},0))
\nonumber\\&+&
(\hat{a}_{1},(\sqrt{\lambda_{x}^{max}},0,0))-(\hat{b}_{1},(0,\sqrt{\lambda_{y}^{max}},0))
\nonumber\\&+&(\hat{b}_{1},(\sqrt{\lambda_{x}^{max}},0,0))]\nonumber\\&=&\|\hat{a}_{1}^{max}\|\|(0,\sqrt{\lambda_{y}^{max}},0)\|
\nonumber\\&+&\|\hat{b}_{1}^{max}\|\|(0,\sqrt{\lambda_{y}^{max}},0)\|\nonumber\\&=&2\sqrt{\lambda_{y}^{max}}
\label{result1}
\end{eqnarray}
where $\hat{a}_{1}^{max}$ is the unit vector parallel to
$(0,\sqrt{\lambda_{y}^{max}},0)$ and perpendicular to
$(\sqrt{\lambda_{x}^{max}},0,0)$; $\hat{b}_{1}^{max}$ is the unit
vectors antiparallel to $(0,\sqrt{\lambda_{y}^{max}},0)$ and
perpendicular to
$(\sqrt{\lambda_{x}^{max}},0,0)$.\\

\textbf{Case-III:} In this case we choose the vectors in such a way
that the maximized  expectation value of the Mermin operator is given by  $\langle B_{M}^{(3)}\rangle_{\rho}=2\sqrt{\lambda_{z}^{max}}$.\\
Again, repeating the above steps, we find
(i) The vector $\hat{a}_{3}^{T}\vec{T}\hat{a}_{2}$ can be
simplified to
\begin{eqnarray}
\hat{a}_{3}^{T}\vec{T}\hat{a}_{2}=(0,0,\sqrt{\lambda_{z}^{max}})
\label{vector111a}
\end{eqnarray},
(ii) The vector $\hat{a}_{3}^{T}\vec{T}\hat{b}_{2}$ can be
simplified to
\begin{eqnarray}
\hat{a}_{3}^{T}\vec{T}\hat{b}_{2}=(-\sqrt{\lambda_{x}^{max}},0,0)
\label{vector222a}
\end{eqnarray}
(iii) The vector $\hat{b}_{3}^{T}\vec{T}\hat{b}_{2}$ can be
simplified to
\begin{eqnarray}
\hat{b}_{3}^{T}\vec{T}\hat{b}_{2}&=&((\hat{b}_{3},T_{x}\hat{b}_{2}),(\hat{b}_{3},T_{y}\hat{b}_{2}),(\hat{b}_{3},T_{z}\hat{b}_{2}))
\nonumber\\&=&(-\|\hat{b}_{3}^{min}\|\|T_{x}\hat{b}_{2}\|,0,0)\nonumber\\&=&(-\sqrt{\lambda_{x}^{max}},0,0)
\label{vector333}
\end{eqnarray}
(iv) The vector $\hat{b}_{3}^{T}\vec{T}\hat{a}_{2}$ can be
simplified to
\begin{eqnarray}
\hat{b}_{3}^{T}\vec{T}\hat{a}_{2}&=&((\hat{b}_{3},T_{x}\hat{a}_{2}),(\hat{b}_{3},T_{y}\hat{a}_{2}),(\hat{b}_{3},T_{z}\hat{a}_{2}))
\nonumber\\&=&(0,0,\|\hat{b}_{3}^{max}\|\|T_{z}\hat{a}_{2}\|)=(0,0,\sqrt{\lambda_{z}^{max}})
\label{vector444}
\end{eqnarray}

Now, from
(\ref{expectationvalue}),(\ref{vector111a}),(\ref{vector222a}),(\ref{vector333}),(\ref{vector444}),
we have
\begin{eqnarray}
\langle
B_{M}^{(3)}\rangle_{\rho}&=&max_{\hat{a}_{1},\hat{b}_{1}}[(\hat{a}_{1},(0,0,\sqrt{\lambda_{z}^{max}}))
\nonumber\\&+&
(\hat{a}_{1},(\sqrt{\lambda_{x}^{max}},0,0))-(\hat{b}_{1},(0,0,\sqrt{\lambda_{z}^{max}}))
\nonumber\\&+&(\hat{b}_{1},(\sqrt{\lambda_{x}^{max}},0,0))]\nonumber\\&=&\|\hat{a}_{1}^{max}\|\|(0,0,\sqrt{\lambda_{z}^{max}})\|
\nonumber\\&+&\|\hat{b}_{1}^{max}\|\|(0,0,\sqrt{\lambda_{z}^{max}})\|\nonumber\\&=&2\sqrt{\lambda_{z}^{max}}
\label{result1}
\end{eqnarray}
where, $\hat{a}_{1}^{max}$ is the unit vector parallel to
$(0,0,\sqrt{\lambda_{z}^{max}})$ and perpendicular to
$(\sqrt{\lambda_{x}^{max}},0,0)$; $\hat{b}_{1}^{max}$ is the unit
vectors antiparallel to $(0,0,\sqrt{\lambda_{z}^{max}})$ and
perpendicular to
$(\sqrt{\lambda_{x}^{max}},0,0)$.

Thus finally, the maximum expectation value of the Mermin operator with
respect to the state $\rho$ is given by
\begin{eqnarray}
\langle B_{M}^{max}\rangle_{\rho}&=&max\{\langle
B_{M}^{(1)}\rangle_{\rho},\langle
B_{M}^{(2)}\rangle_{\rho},\langle
B_{M}^{(3)}\rangle_{\rho}\}\nonumber\\
&=&max\{2\sqrt{\lambda_{x}^{max}},2\sqrt{\lambda_{y}^{max}}
,2\sqrt{\lambda_{z}^{max}}\}
\end{eqnarray}
The Mermin inequality is violated if
\begin{eqnarray}
&&\langle B_{M}^{max}\rangle_{\rho}>2\nonumber\\&&\Rightarrow
max\{2\sqrt{\lambda_{x}^{max}},2\sqrt{\lambda_{y}^{max}},2\sqrt{\lambda_{z}^{max}}\}>2
\nonumber\\&&\Rightarrow
max\{\sqrt{\lambda_{x}^{max}},\sqrt{\lambda_{y}^{max}}
,\sqrt{\lambda_{z}^{max}}\}>1
\end{eqnarray}
Hence, proved.


\textbf{Theorem-2:} If the symmetric matrices
$T_{x}^{T}T_{x}$,$T_{y}^{T}T_{y}$ and $T_{z}^{T}T_{z}$ have two
equal largest eigenvalue $\lambda_{x}^{max}$,$\lambda_{y}^{max}$
and $\lambda_{z}^{max}$ respectively then Mermin's inequality is
violated if
\begin{eqnarray}
\langle B_{M}^{max}\rangle_{\rho}&=&max_{B_{M}}Tr(B_{M}\rho)
\nonumber\\&=&max\{4\sqrt{\lambda_{x}^{max}},4\sqrt{\lambda_{y}^{max}},4\sqrt{\lambda_{z}^{max}}\}\nonumber\\&>&2\label{theorem2}
\end{eqnarray}
\textbf{Proof:} In the expression for $\langle
B_{M}\rangle_{\rho}$ given by Eq. (\ref{expectationvalue}), we
first simplify the vectors $\hat{a}_{3}^{T}\vec{T}\hat{a}_{2}$,
$\hat{b}_{3}^{T}\vec{T}\hat{b}_{2}$,
$\hat{b}_{3}^{T}\vec{T}\hat{a}_{2}$,
$\hat{a}_{3}^{T}\vec{T}\hat{b}_{2}$. We again consider the following cases:\\
\textbf{Case-I:} We consider the symmetric matrix
$T_{x}^{T}T_{x}$ which has two equal largest eigenvalues
$\lambda_{x}^{max}$ and choose the unit vectors in such a way that
it maximizes the expectation value of the Mermin operator given by
 $\langle B_{M}^{(4)}\rangle_{\rho}=4\sqrt{\lambda_{x}^{max}}$.\\
(i) The vector $\hat{a}_{3}^{T}\vec{T}\hat{a}_{2}$ can be
simplified as
\begin{eqnarray}
\hat{a}_{3}^{T}\vec{T}\hat{a}_{2}&=&((\hat{a}_{3},T_{x}\hat{a}_{2}),(\hat{a}_{3},T_{y}\hat{a}_{2}),(\hat{a}_{3},T_{z}\hat{a}_{2}))
\nonumber\\&=&(\|\hat{a}_{3}^{max}\|\|T_{x}\hat{a}_{2}\|,0,0)
\label{vector5}
\end{eqnarray}
where $\hat{a}_{3}^{max}$ is the unit vector along
$T_{x}\hat{a}_{2}$ and perpendicular to $T_{y}\hat{a}_{2}$ and
$T_{z}\hat{a}_{2}$.
Since $\hat{a}_{3}^{max}$ is a unit vector,
$\|\hat{a}_{3}^{max}\|=1$. Again,
$\|T_{x}\hat{a}_{2}\|^{2}=(T_{x}\hat{a}_{2},T_{x}\hat{a}_{2})=(\hat{a}_{2},T_{x}^{T}T_{x}\hat{a}_{2})$.
If $\lambda_{x}^{max}$ is the largest eigenvalue of the symmetric
matrix $T_{x}^{T}T_{x}$ and $\hat{a}_{2}^{max}$ is the
corresponding unit eigenvector then
$\|T_{x}\hat{a}_{2}\|^{2}=(\hat{a}_{2},T_{x}^{T}T_{x}\hat{a}_{2})=(\hat{a}_{2},\lambda_{x}^{max}\hat{a}_{2}^{max})$.
If $\hat{a}_{2}$ is the unit vector along $\hat{a}_{2}^{max}$ then
$\|T_{x}\hat{a}_{2}\|^{2}=\lambda_{x}^{max}$. Thus,
\begin{eqnarray}
\hat{a}_{3}^{T}\vec{T}\hat{a}_{2}=(\sqrt{\lambda_{x}^{max}},0,0)
\label{vector5a}
\end{eqnarray}
(ii) The vector $\hat{a}_{3}^{T}\vec{T}\hat{b}_{2}$ can be
simplified as
\begin{eqnarray}
\hat{a}_{3}^{T}\vec{T}\hat{b}_{2}&=&((\hat{a}_{3},T_{x}\hat{b}_{2}),(\hat{a}_{3},T_{y}\hat{b}_{2}),(\hat{a}_{3},T_{z}\hat{b}_{2}))
\nonumber\\&=&(-\|\hat{a}_{3}^{min}\|\|T_{x}\hat{b}_{2}\|,0,0)
\label{vector6}
\end{eqnarray}
where $\hat{a}_{3}^{min}$ is the unit vector antiparallel to
$T_{x}\hat{b}_{2}$ and perpendicular to $T_{y}\hat{b}_{2}$ and
$T_{z}\hat{b}_{2}$. Since $\hat{a}_{3}^{min}$ is the unit vector
so $\|\hat{a}_{3}^{min}\|=1$. Again,
$\|T_{x}\hat{b}_{2}\|^{2}=(T_{x}\hat{b}_{2},T_{x}\hat{b}_{2})=(\hat{b}_{2},T_{x}^{T}T_{x}\hat{b}_{2})$.
Since the matrix $T_{x}^{T}T_{x}$ has two equal largest eigenvalues
$\lambda_{x}^{max}$, so $\hat{b}_{2}^{max}$ is another
corresponding unit eigenvector. Then
$\|T_{x}\hat{b}_{2}\|^{2}=(\hat{b}_{2},T_{x}^{T}T_{x}\hat{b}_{2})=(\hat{b}_{2},\lambda_{x}^{max}\hat{b}_{2}^{max})$.
If $\hat{b}_{2}$ is the unit vector along $\hat{b}_{2}^{max}$ then
$\|T_{x}\hat{b}_{2}\|^{2}=\lambda_{x}^{max}$. Thus,
\begin{eqnarray}
\hat{a}_{3}^{T}\vec{T}\hat{b}_{2}=(-\sqrt{\lambda_{x}^{max}},0,0)
\label{vector6a}
\end{eqnarray}
(iii) The vector $\hat{b}_{3}^{T}\vec{T}\hat{b}_{2}$ can be
simplified as
\begin{eqnarray}
\hat{b}_{3}^{T}\vec{T}\hat{b}_{2}&=&((\hat{b}_{3},T_{x}\hat{b}_{2}),(\hat{b}_{3},T_{y}\hat{b}_{2}),(\hat{b}_{3},T_{z}\hat{b}_{2}))
\nonumber\\&=&(-\|\hat{b}_{3}^{min}\|\|T_{x}\hat{b}_{2}\|,0,0)\nonumber\\&=&(-\sqrt{\lambda_{x}^{max}},0,0)
\label{vector7}
\end{eqnarray}
where $\hat{b}_{3}^{min}$ is the unit vector antiparallel to
$T_{x}\hat{b}_{2}$ and perpendicular to $T_{y}\hat{b}_{2}$ and
$T_{z}\hat{b}_{2}$.\\
(iv) The vector $\hat{b}_{3}^{T}\vec{T}\hat{a}_{2}$ can be
simplified as
\begin{eqnarray}
\hat{b}_{3}^{T}\vec{T}\hat{a}_{2}&=&((\hat{b}_{3},T_{x}\hat{a}_{2}),(\hat{b}_{3},T_{y}\hat{a}_{2}),(\hat{b}_{3},T_{z}\hat{a}_{2}))
\nonumber\\&=&(-\|\hat{b}_{3}^{min}\|\|T_{x}\hat{a}_{2}\|,0,0)\nonumber\\&=&(-\sqrt{\lambda_{x}^{max}},0,0)
\label{vector8}
\end{eqnarray}
where $\hat{b}_{3}^{min}$ is the unit vector antiparallel to
$T_{x}\hat{a}_{2}$ and perpendicular to $T_{y}\hat{a}_{2}$ and
$T_{z}\hat{a}_{2}$.\\

From
(\ref{expectationvalue}),(\ref{vector5a}),(\ref{vector6a}),(\ref{vector7}),(\ref{vector8}),
we have
\begin{eqnarray}
\langle
B_{M}^{(4)}\rangle_{\rho}&=&max_{\hat{a}_{1},\hat{b}_{1}}[(\hat{a}_{1},(\sqrt{\lambda_{x}^{max}},0,0))
\nonumber\\&+&
(\hat{a}_{1},(\sqrt{\lambda_{x}^{max}},0,0))+(\hat{b}_{1},(\sqrt{\lambda_{x}^{max}},0,0))
\nonumber\\&+&(\hat{b}_{1},(\sqrt{\lambda_{x}^{max}},0,0))]\nonumber\\&=&2\|\hat{a}_{1}^{max}\|
\|(\sqrt{\lambda_{x}^{max}},0,0))\|\nonumber\\&+&2\|\hat{b}_{1}^{max}\|\|(\sqrt{\lambda_{x}^{max}},0,0))\|\nonumber\\&=&
4\sqrt{\lambda_{x}^{max}} \label{result44}
\end{eqnarray}
where $\hat{a}_{1}^{max}$ and $\hat{b}_{1}^{max}$ is the unit vector along $(\sqrt{\lambda_{x}^{max}},0,0)$.\\

\textbf{Case-II:} Here we consider the symmetric matrix
$T_{y}^{T}T_{y}$ which has two equal largest eigenvalues
$\lambda_{y}^{max}$ and choose the unit vectors in such a way that
it maximizes the expectation value of the Mermin operator given by
 $\langle B_{M}^{(5)}\rangle_{\rho}=4\sqrt{\lambda_{y}^{max}}$.\\
(i) The vector $\hat{a}_{3}^{T}\vec{T}\hat{a}_{2}$ can be
simplified as
\begin{eqnarray}
\hat{a}_{3}^{T}\vec{T}\hat{a}_{2}&=&((\hat{a}_{3},T_{x}\hat{a}_{2}),(\hat{a}_{3},T_{y}\hat{a}_{2}),(\hat{a}_{3},T_{z}\hat{a}_{2}))
\nonumber\\&=&(0,\|\hat{a}_{3}^{max}\|\|T_{y}\hat{a}_{2}\|,0)
\label{vector55}
\end{eqnarray}
where $\hat{a}_{3}^{max}$ is the unit vector along
$T_{y}\hat{a}_{2}$ and perpendicular to $T_{x}\hat{a}_{2}$ and
$T_{z}\hat{a}_{2}$.
Since $\hat{a}_{3}^{max}$ is a unit vector,
$\|\hat{a}_{3}^{max}\|=1$. Again,
$\|T_{y}\hat{a}_{2}\|^{2}=(T_{y}\hat{a}_{2},T_{y}\hat{a}_{2})=(\hat{a}_{2},T_{y}^{T}T_{y}\hat{a}_{2})$.
If $\lambda_{y}^{max}$ is the largest eigenvalue of the symmetric
matrix $T_{y}^{T}T_{y}$ and $\hat{a}_{2}^{max}$ is the
corresponding unit eigenvector then
$\|T_{y}\hat{a}_{2}\|^{2}=(\hat{a}_{2},T_{y}^{T}T_{y}\hat{a}_{2})=(\hat{a}_{2},\lambda_{y}^{max}\hat{a}_{2}^{max})$.
If $\hat{a}_{2}$ is the unit vector along $\hat{a}_{2}^{max}$ then
$\|T_{y}\hat{a}_{2}\|^{2}=\lambda_{y}^{max}$. Thus,
\begin{eqnarray}
\hat{a}_{3}^{T}\vec{T}\hat{a}_{2}=(0,\sqrt{\lambda_{y}^{max}},0)
\label{vector55a}
\end{eqnarray}
(ii) The vector $\hat{a}_{3}^{T}\vec{T}\hat{b}_{2}$ can be
simplified as
\begin{eqnarray}
\hat{a}_{3}^{T}\vec{T}\hat{b}_{2}&=&((\hat{a}_{3},T_{x}\hat{b}_{2}),(\hat{a}_{3},T_{y}\hat{b}_{2}),(\hat{a}_{3},T_{z}\hat{b}_{2}))
\nonumber\\&=&(0,-\|\hat{a}_{3}^{min}\|\|T_{y}\hat{b}_{2}\|,0)
\label{vector66}
\end{eqnarray}
where $\hat{a}_{3}^{min}$ is the unit vector antiparallel to
$T_{y}\hat{b}_{2}$ and perpendicular to $T_{x}\hat{b}_{2}$ and
$T_{z}\hat{b}_{2}$. Since $\hat{a}_{3}^{min}$ is a unit vector
so $\|\hat{a}_{3}^{min}\|=1$. Again,
$\|T_{y}\hat{b}_{2}\|^{2}=(T_{y}\hat{b}_{2},T_{y}\hat{b}_{2})=(\hat{b}_{2},T_{y}^{T}T_{y}\hat{b}_{2})$.
Since the matrix $T_{y}^{T}T_{y}$ has two equal largest eigenvalues
$\lambda_{y}^{max}$, so let us consider $\hat{b}_{2}^{max}$ be
another corresponding unit eigenvector. Then
$\|T_{y}\hat{b}_{2}\|^{2}=(\hat{b}_{2},T_{y}^{T}T_{y}\hat{b}_{2})=(\hat{b}_{2},\lambda_{y}^{max}\hat{b}_{2}^{max})$.
If $\hat{b}_{2}$ is the unit vector along $\hat{b}_{2}^{max}$, then
$\|T_{y}\hat{b}_{2}\|^{2}=\lambda_{y}^{max}$. Thus,
\begin{eqnarray}
\hat{a}_{3}^{T}\vec{T}\hat{b}_{2}=(0,-\sqrt{\lambda_{y}^{max}},0)
\label{vector66a}
\end{eqnarray}
(iii) The vector $\hat{b}_{3}^{T}\vec{T}\hat{b}_{2}$ can be
simplified as
\begin{eqnarray}
\hat{b}_{3}^{T}\vec{T}\hat{b}_{2}&=&((\hat{b}_{3},T_{x}\hat{b}_{2}),(\hat{b}_{3},T_{y}\hat{b}_{2}),(\hat{b}_{3},T_{z}\hat{b}_{2}))
\nonumber\\&=&(0,-\|\hat{b}_{3}^{min}\|\|T_{y}\hat{b}_{2}\|,0)\nonumber\\&=&(0,-\sqrt{\lambda_{y}^{max}},0)
\label{vector77}
\end{eqnarray}
where $\hat{b}_{3}^{min}$ is the unit vector antiparallel to
$T_{y}\hat{b}_{2}$ and perpendicular to $T_{x}\hat{b}_{2}$ and
$T_{z}\hat{b}_{2}$.\\
(iv) The vector $\hat{b}_{3}^{T}\vec{T}\hat{a}_{2}$ can be
simplified as
\begin{eqnarray}
\hat{b}_{3}^{T}\vec{T}\hat{a}_{2}&=&((\hat{b}_{3},T_{x}\hat{a}_{2}),(\hat{b}_{3},T_{y}\hat{a}_{2}),(\hat{b}_{3},T_{z}\hat{a}_{2}))
\nonumber\\&=&(0,-\|\hat{b}_{3}^{min}\|\|T_{y}\hat{a}_{2}\|,0)\nonumber\\&=&(0,-\sqrt{\lambda_{y}^{max}},0)
\label{vector88}
\end{eqnarray}
where $\hat{b}_{3}^{min}$ is the unit vector antiparallel to
$T_{y}\hat{a}_{2}$ and perpendicular to $T_{x}\hat{a}_{2}$ and
$T_{z}\hat{a}_{2}$.\\

From
(\ref{expectationvalue}),(\ref{vector55a}),(\ref{vector66a}),(\ref{vector77}),(\ref{vector88}),
we have
\begin{eqnarray}
\langle
B_{M}^{(5)}\rangle_{\rho}&=&max_{\hat{a}_{1},\hat{b}_{1}}[(\hat{a}_{1},(0,\sqrt{\lambda_{y}^{max}},0))
\nonumber\\&+&
(\hat{a}_{1},(0,\sqrt{\lambda_{y}^{max}},0))+(\hat{b}_{1},(0,\sqrt{\lambda_{y}^{max}},0))
\nonumber\\&+&(\hat{b}_{1},(0,\sqrt{\lambda_{y}^{max}},0))]\nonumber\\&=&2\|\hat{a}_{1}^{max}\|
\|(0,\sqrt{\lambda_{y}^{max}},0))\|\nonumber\\&+&2\|\hat{b}_{1}^{max}\|\|(0,\sqrt{\lambda_{y}^{max}},0))\|\nonumber\\&=&
4\sqrt{\lambda_{y}^{max}} \label{result55}
\end{eqnarray}
where $\hat{a}_{1}^{max}$ and $\hat{b}_{1}^{max}$ is the unit vector along $(0,\sqrt{\lambda_{y}^{max}},0)$.\\

\textbf{Case-III:} Here we consider the symmetric matrix
$T_{y}^{T}T_{y}$ which has two equal largest eigenvalues
$\lambda_{y}^{max}$ and choose the unit vectors in such a way that
it maximizes the expectation value of the Mermin operator, given by
$\langle B_{M}^{(6)}\rangle_{\rho}=4\sqrt{\lambda_{z}^{max}}$.\\
(i) The vector $\hat{a}_{3}^{T}\vec{T}\hat{a}_{2}$ can be
simplified as
\begin{eqnarray}
\hat{a}_{3}^{T}\vec{T}\hat{a}_{2}&=&((\hat{a}_{3},T_{x}\hat{a}_{2}),(\hat{a}_{3},T_{y}\hat{a}_{2}),(\hat{a}_{3},T_{z}\hat{a}_{2}))
\nonumber\\&=&(0,0,\|\hat{a}_{3}^{max}\|\|T_{z}\hat{a}_{2}\|)
\label{vector555}
\end{eqnarray}
where $\hat{a}_{3}^{max}$ is the unit vector along
$T_{z}\hat{a}_{2}$ and perpendicular to $T_{x}\hat{a}_{2}$ and
$T_{y}\hat{a}_{2}$.
Since $\hat{a}_{3}^{max}$ is a unit vector,
$\|\hat{a}_{3}^{max}\|=1$. Again,
$\|T_{z}\hat{a}_{2}\|^{2}=(T_{z}\hat{a}_{2},T_{z}\hat{a}_{2})=(\hat{a}_{2},T_{z}^{T}T_{z}\hat{a}_{2})$.
If $\lambda_{z}^{max}$ is the largest eigenvalue of the symmetric
matrix $T_{z}^{T}T_{z}$ and $\hat{a}_{2}^{max}$ is the
corresponding unit eigenvector, then
$\|T_{z}\hat{a}_{2}\|^{2}=(\hat{a}_{2},T_{z}^{T}T_{z}\hat{a}_{2})=(\hat{a}_{2},\lambda_{z}^{max}\hat{a}_{2}^{max})$.
If $\hat{a}_{2}$ is the unit vector along $\hat{a}_{2}^{max}$, then
$\|T_{z}\hat{a}_{2}\|^{2}=\lambda_{z}^{max}$. Thus,
\begin{eqnarray}
\hat{a}_{3}^{T}\vec{T}\hat{a}_{2}=(0,0,\sqrt{\lambda_{z}^{max}})
\label{vector555a}
\end{eqnarray}
(ii) The vector $\hat{a}_{3}^{T}\vec{T}\hat{b}_{2}$ can be
simplified as
\begin{eqnarray}
\hat{a}_{3}^{T}\vec{T}\hat{b}_{2}&=&((\hat{a}_{3},T_{x}\hat{b}_{2}),(\hat{a}_{3},T_{y}\hat{b}_{2}),(\hat{a}_{3},T_{z}\hat{b}_{2}))
\nonumber\\&=&(0,0,-\|\hat{a}_{3}^{min}\|\|T_{z}\hat{b}_{2}\|)
\label{vector666}
\end{eqnarray}
where $\hat{a}_{3}^{min}$ is the unit vector antiparallel to
$T_{z}\hat{b}_{2}$ and perpendicular to $T_{x}\hat{b}_{2}$ and
$T_{y}\hat{b}_{2}$. Since $\hat{a}_{3}^{min}$ is a unit vector,
 $\|\hat{a}_{3}^{min}\|=1$. Again,
$\|T_{z}\hat{b}_{2}\|^{2}=(T_{z}\hat{b}_{2},T_{z}\hat{b}_{2})=(\hat{b}_{2},T_{z}^{T}T_{z}\hat{b}_{2})$.
Since the matrix $T_{z}^{T}T_{z}$ has two equal largest eigenvalues
$\lambda_{z}^{max}$, let us consider $\hat{b}_{2}^{max}$ to be
another corresponding unit eigenvector. Then
$\|T_{z}\hat{b}_{2}\|^{2}=(\hat{b}_{2},T_{z}^{T}T_{z}\hat{b}_{2})=(\hat{b}_{2},\lambda_{z}^{max}\hat{b}_{2}^{max})$.
If $\hat{b}_{2}$ is the unit vector along $\hat{b}_{2}^{max}$, then
$\|T_{z}\hat{b}_{2}\|^{2}=\lambda_{z}^{max}$. Thus,
\begin{eqnarray}
\hat{a}_{3}^{T}\vec{T}\hat{b}_{2}=(0,0,-\sqrt{\lambda_{z}^{max}})
\label{vector666a}
\end{eqnarray}
(iii) The vector $\hat{b}_{3}^{T}\vec{T}\hat{b}_{2}$ can be
simplified as
\begin{eqnarray}
\hat{b}_{3}^{T}\vec{T}\hat{b}_{2}&=&((\hat{b}_{3},T_{x}\hat{b}_{2}),(\hat{b}_{3},T_{y}\hat{b}_{2}),(\hat{b}_{3},T_{z}\hat{b}_{2}))
\nonumber\\&=&(0,0,-\|\hat{b}_{3}^{min}\|\|T_{y}\hat{b}_{2}\|)\nonumber\\&=&(0,0,-\sqrt{\lambda_{z}^{max}})
\label{vector777}
\end{eqnarray}
where $\hat{b}_{3}^{min}$ is the unit vector antiparallel to
$T_{z}\hat{b}_{2}$ and perpendicular to $T_{x}\hat{b}_{2}$ and
$T_{y}\hat{b}_{2}$.\\
(iv) The vector $\hat{b}_{3}^{T}\vec{T}\hat{a}_{2}$ can be
simplified as
\begin{eqnarray}
\hat{b}_{3}^{T}\vec{T}\hat{a}_{2}&=&((\hat{b}_{3},T_{x}\hat{a}_{2}),(\hat{b}_{3},T_{y}\hat{a}_{2}),(\hat{b}_{3},T_{z}\hat{a}_{2}))
\nonumber\\&=&(0,0,-\|\hat{b}_{3}^{min}\|\|T_{y}\hat{a}_{2}\|)\nonumber\\&=&(0,0,-\sqrt{\lambda_{z}^{max}})
\label{vector888}
\end{eqnarray}
where $\hat{b}_{3}^{min}$ is the unit vector antiparallel to
$T_{z}\hat{a}_{2}$ and perpendicular to $T_{x}\hat{a}_{2}$ and
$T_{y}\hat{a}_{2}$.\\

From
(\ref{expectationvalue}),(\ref{vector555a}),(\ref{vector666a}),(\ref{vector777}),(\ref{vector888}),
we have
\begin{eqnarray}
\langle
B_{M}^{(6)}\rangle_{\rho}&=&max_{\hat{a}_{1},\hat{b}_{1}}[(\hat{a}_{1},(0,0,\sqrt{\lambda_{z}^{max}}))
\nonumber\\&+&
(\hat{a}_{1},(0,0,\sqrt{\lambda_{z}^{max}}))+(\hat{b}_{1},(0,0,\sqrt{\lambda_{z}^{max}}))
\nonumber\\&+&(\hat{b}_{1},(0,0,\sqrt{\lambda_{z}^{max}}))]\nonumber\\&=&2\|\hat{a}_{1}^{max}\|
\|(0,0,\sqrt{\lambda_{z}^{max}}))\|\nonumber\\&+&2\|\hat{b}_{1}^{max}\|\|(0,0,\sqrt{\lambda_{z}^{max}}))\|\nonumber\\&=&
4\sqrt{\lambda_{z}^{max}} \label{result66}
\end{eqnarray}
where $\hat{a}_{1}^{max}$ and $\hat{b}_{1}^{max}$ is the unit vector along $(0,0,\sqrt{\lambda_{z}^{max}})$.\\
Thus, the maximum expectation value of the Mermin operator with
respect to the state $\rho$ is given by
\begin{eqnarray}
\langle B_{M}^{max}\rangle_{\rho}&=&max\{\langle
B_{M}^{(4)}\rangle_{\rho},\langle
B_{M}^{(5)}\rangle_{\rho},\langle
B_{M}^{(6)}\rangle_{\rho}\}\nonumber\\
&=&max\{4\sqrt{\lambda_{x}^{max}},4\sqrt{\lambda_{y}^{max}}
,4\sqrt{\lambda_{z}^{max}}\}
\end{eqnarray}
The Mermin inequality is violated if
\begin{eqnarray}
\langle
B_{M}^{max}\rangle_{\rho}=max\{4\sqrt{\lambda_{x}^{max}},4\sqrt{\lambda_{y}^{max}},4\sqrt{\lambda_{z}^{max}}\}>2
\end{eqnarray}
Thus
$max\{\sqrt{\lambda_{x}^{max}},\sqrt{\lambda_{y}^{max}},\sqrt{\lambda_{z}^{max}}\}>\frac{1}{2}$.
Hence, proved.\\

\textbf{Theorem-3:} If $T_{x}^{T}T_{x}$ has two equal largest
eigenvalues $\lambda_{x}^{max}$ and $T_{y}^{T}T_{y}$ and
$T_{z}^{T}T_{z}$ has unique largest eigenvalue $\lambda_{y}^{max}$
and $\lambda_{z}^{max}$ respectively, then Mermin's inequality is
violated if
\begin{eqnarray}
\langle
B_{M}^{max}\rangle_{\rho}&=&max\{4\sqrt{\lambda_{x}^{max}},2\sqrt{\lambda_{y}^{max}}
,2\sqrt{\lambda_{z}^{max}}\}\nonumber\\&>&2\label{theorem3}
\end{eqnarray}
\textbf{Proof:} In the expression for $\langle
B_{M}\rangle_{\rho}$ given by Eq. (\ref{expectationvalue}), we
first simplify the vectors $\hat{a}_{3}^{T}\vec{T}\hat{a}_{2}$,
$\hat{b}_{3}^{T}\vec{T}\hat{b}_{2}$,
$\hat{b}_{3}^{T}\vec{T}\hat{a}_{2}$,
$\hat{a}_{3}^{T}\vec{T}\hat{b}_{2}$. \\
\textbf{Case-I:} In this case we choose the vectors in such a way
that it maximizes the expectation value of the Mermin operator, given by  $\langle B_{M}^{(7)}\rangle_{\rho}=4\sqrt{\lambda_{x}^{max}}$.\\
(i) The vector $\hat{a}_{3}^{T}\vec{T}\hat{a}_{2}$ can be
simplified as
\begin{eqnarray}
\hat{a}_{3}^{T}\vec{T}\hat{a}_{2}&=&((\hat{a}_{3},T_{x}\hat{a}_{2}),(\hat{a}_{3},T_{y}\hat{a}_{2}),(\hat{a}_{3},T_{z}\hat{a}_{2}))
\nonumber\\&=&(\|\hat{a}_{3}^{max}\|\|T_{x}\hat{a}_{2}\|,0,0)
\label{vector35}
\end{eqnarray}
where $\hat{a}_{3}^{max}$ is the unit vector along
$T_{x}\hat{a}_{2}$ and perpendicular to $T_{y}\hat{a}_{2}$ and
$T_{z}\hat{a}_{2}$.
Since $\hat{a}_{3}^{max}$ is a unit vector,
$\|\hat{a}_{3}^{max}\|=1$. Again,
$\|T_{x}\hat{a}_{2}\|^{2}=(T_{x}\hat{a}_{2},T_{x}\hat{a}_{2})=(\hat{a}_{2},T_{x}^{T}T_{x}\hat{a}_{2})$.
If $\lambda_{x}^{max}$ is the largest eigenvalue of the symmetric
matrix $T_{x}^{T}T_{x}$ and $\hat{a}_{2}^{max}$ is the
corresponding unit eigenvector, then
$\|T_{x}\hat{a}_{2}\|^{2}=(\hat{a}_{2},T_{x}^{T}T_{x}\hat{a}_{2})=(\hat{a}_{2},\lambda_{x}^{max}\hat{a}_{2}^{max})$.
If $\hat{a}_{2}$ is the unit vector along $\hat{a}_{2}^{max}$, then
$\|T_{x}\hat{a}_{2}\|^{2}=\lambda_{x}^{max}$. Thus,
\begin{eqnarray}
\hat{a}_{3}^{T}\vec{T}\hat{a}_{2}=(\sqrt{\lambda_{x}^{max}},0,0)
\label{vector35a}
\end{eqnarray}
(ii) The vector $\hat{a}_{3}^{T}\vec{T}\hat{b}_{2}$ can be
simplified as
\begin{eqnarray}
\hat{a}_{3}^{T}\vec{T}\hat{b}_{2}&=&((\hat{a}_{3},T_{x}\hat{b}_{2}),(\hat{a}_{3},T_{y}\hat{b}_{2}),(\hat{a}_{3},T_{z}\hat{b}_{2}))
\nonumber\\&=&(-\|\hat{a}_{3}^{min}\|\|T_{x}\hat{b}_{2}\|,0,0)
\label{vector36}
\end{eqnarray}
where $\hat{a}_{3}^{min}$ is the unit vector antiparallel to
$T_{x}\hat{b}_{2}$ and perpendicular to $T_{y}\hat{b}_{2}$ and
$T_{z}\hat{b}_{2}$. Since $\hat{a}_{3}^{min}$ is a unit vector,
 $\|\hat{a}_{3}^{min}\|=1$. Again,
$\|T_{x}\hat{b}_{2}\|^{2}=(T_{x}\hat{b}_{2},T_{x}\hat{b}_{2})=(\hat{b}_{2},T_{x}^{T}T_{x}\hat{b}_{2})$.
Since the matrix $T_{x}^{T}T_{x}$ has two equal largest eigenvalues
$\lambda_{x}^{max}$, so $\hat{b}_{2}^{max}$ is another
corresponding unit eigenvector. Then
$\|T_{x}\hat{b}_{2}\|^{2}=(\hat{b}_{2},T_{x}^{T}T_{x}\hat{b}_{2})=(\hat{b}_{2},\lambda_{x}^{max}\hat{b}_{2}^{max})$.
If $\hat{b}_{2}$ is the unit vector along $\hat{b}_{2}^{max}$, then
$\|T_{x}\hat{b}_{2}\|^{2}=\lambda_{x}^{max}$. Thus,
\begin{eqnarray}
\hat{a}_{3}^{T}\vec{T}\hat{b}_{2}=(-\sqrt{\lambda_{x}^{max}},0,0)
\label{vector36a}
\end{eqnarray}
(iii) The vector $\hat{b}_{3}^{T}\vec{T}\hat{b}_{2}$ can be
simplified as
\begin{eqnarray}
\hat{b}_{3}^{T}\vec{T}\hat{b}_{2}&=&((\hat{b}_{3},T_{x}\hat{b}_{2}),(\hat{b}_{3},T_{y}\hat{b}_{2}),(\hat{b}_{3},T_{z}\hat{b}_{2}))
\nonumber\\&=&(-\|\hat{b}_{3}^{min}\|\|T_{x}\hat{b}_{2}\|,0,0)\nonumber\\&=&(-\sqrt{\lambda_{x}^{max}},0,0)
\label{vector37}
\end{eqnarray}
where $\hat{b}_{3}^{min}$ is the unit vector antiparallel to
$T_{x}\hat{b}_{2}$ and perpendicular to $T_{y}\hat{b}_{2}$ and
$T_{z}\hat{b}_{2}$.\\
(iv) The vector $\hat{b}_{3}^{T}\vec{T}\hat{a}_{2}$ can be
simplified as
\begin{eqnarray}
\hat{b}_{3}^{T}\vec{T}\hat{a}_{2}&=&((\hat{b}_{3},T_{x}\hat{a}_{2}),(\hat{b}_{3},T_{y}\hat{a}_{2}),(\hat{b}_{3},T_{z}\hat{a}_{2}))
\nonumber\\&=&(-\|\hat{b}_{3}^{min}\|\|T_{x}\hat{a}_{2}\|,0,0)\nonumber\\&=&(-\sqrt{\lambda_{x}^{max}},0,0)
\label{vector38}
\end{eqnarray}
where $\hat{b}_{3}^{min}$ is the unit vector antiparallel to
$T_{x}\hat{a}_{2}$ and perpendicular to $T_{y}\hat{a}_{2}$ and
$T_{z}\hat{a}_{2}$.\\

From
(\ref{expectationvalue}),(\ref{vector35a}),(\ref{vector36a}),(\ref{vector37}),(\ref{vector38}),
we have
\begin{eqnarray}
\langle
B_{M}^{(7)}\rangle_{\rho}&=&max_{\hat{a}_{1},\hat{b}_{1}}[(\hat{a}_{1},(\sqrt{\lambda_{x}^{max}},0,0))
\nonumber\\&+&
(\hat{a}_{1},(\sqrt{\lambda_{x}^{max}},0,0))+(\hat{b}_{1},(\sqrt{\lambda_{x}^{max}},0,0))
\nonumber\\&+&(\hat{b}_{1},(\sqrt{\lambda_{x}^{max}},0,0))]\nonumber\\&=&2\|\hat{a}_{1}^{max}\|
\|(\sqrt{\lambda_{x}^{max}},0,0))\|\nonumber\\&+&2\|\hat{b}_{1}^{max}\|\|(\sqrt{\lambda_{x}^{max}},0,0))\|\nonumber\\&=&
4\sqrt{\lambda_{x}^{max}} \label{result344}
\end{eqnarray}
where $\hat{a}_{1}^{max}$ and $\hat{b}_{1}^{max}$ is the unit vector along $(\sqrt{\lambda_{x}^{max}},0,0)$.\\

\textbf{Case-II:} In this case we choose the vectors in such a way
that it maximizes the expectation value of the Mermin operator, given by  $\langle B_{M}^{(8)}\rangle_{\rho}=2\sqrt{\lambda_{y}^{max}}$.\\
(i) The vector $\hat{a}_{3}^{T}\vec{T}\hat{a}_{2}$ can be
simplified as
\begin{eqnarray}
\hat{a}_{3}^{T}\vec{T}\hat{a}_{2}&=&((\hat{a}_{3},T_{x}\hat{a}_{2}),(\hat{a}_{3},T_{y}\hat{a}_{2}),(\hat{a}_{3},T_{z}\hat{a}_{2}))
\nonumber\\&=&(0,\|\hat{a}_{3}^{max}\|\|T_{y}\hat{a}_{2}\|,0)
\label{vector311}
\end{eqnarray}
where $\hat{a}_{3}^{max}$ is the unit vector along
$T_{y}\hat{a}_{2}$ and perpendicular to $T_{x}\hat{a}_{2}$ and
$T_{z}\hat{a}_{2}$.
Since $\hat{a}_{3}^{max}$ is a unit vector,
$\|\hat{a}_{3}^{max}\|=1$. Again,
$\|T_{y}\hat{a}_{2}\|^{2}=(T_{y}\hat{a}_{2},T_{y}\hat{a}_{2})=(\hat{a}_{2},T_{y}^{T}T_{y}\hat{a}_{2})$.
If $\lambda_{y}^{max}$ is the largest eigenvalue of the symmetric
matrix $T_{y}^{T}T_{y}$ and $\hat{a}_{2}^{max}$ is the
corresponding unit vector then
$\|T_{y}\hat{a}_{2}\|^{2}=(\hat{a}_{2},T_{y}^{T}T_{y}\hat{a}_{2})=(\hat{a}_{2},\lambda_{y}^{max}\hat{a}_{2}^{max})$.
If $\hat{a}_{2}$ is the unit vector along $\hat{a}_{2}^{max}$, then
$\|T_{y}\hat{a}_{2}\|^{2}=\lambda_{y}^{max}$. Thus,
\begin{eqnarray}
\hat{a}_{3}^{T}\vec{T}\hat{a}_{2}=(0,\sqrt{\lambda_{y}^{max}},0)
\label{vector311a}
\end{eqnarray}
(ii) The vector $\hat{a}_{3}^{T}\vec{T}\hat{b}_{2}$ can be
simplified as
\begin{eqnarray}
\hat{a}_{3}^{T}\vec{T}\hat{b}_{2}&=&((\hat{a}_{3},T_{x}\hat{b}_{2}),(\hat{a}_{3},T_{y}\hat{b}_{2}),(\hat{a}_{3},T_{z}\hat{b}_{2}))
\nonumber\\&=&(-\|\hat{a}_{3}^{min}\|\|T_{x}\hat{b}_{2}\|,0,0)
\label{vector322}
\end{eqnarray}
where $\hat{a}_{3}^{min}$ is the unit vector antiparallel to
$T_{x}\hat{b}_{2}$ and perpendicular to $T_{y}\hat{b}_{2}$ and
$T_{z}\hat{b}_{2}$.
Since $\hat{a}_{3}^{min}$ is a unit vector,
$\|\hat{a}_{3}^{min}\|=1$. Again,
$\|T_{x}\hat{b}_{2}\|^{2}=(T_{x}\hat{b}_{2},T_{x}\hat{b}_{2})=(\hat{b}_{2},T_{x}^{T}T_{x}\hat{b}_{2})$.
If $\lambda_{x}^{max}$ is the largest eigenvalue of the symmetric
matrix $T_{x}^{T}T_{x}$ and $\hat{b}_{2}^{max}$ is the
corresponding unit vector, then
$\|T_{x}\hat{b}_{2}\|^{2}=(\hat{b}_{2},T_{x}^{T}T_{x}\hat{b}_{2})=(\hat{b}_{2},\lambda_{x}^{max}\hat{b}_{2}^{max})$.
If $\hat{b}_{2}$ is the unit vector along $\hat{b}_{2}^{max}$, then
$\|T_{x}\hat{b}_{2}\|^{2}=\lambda_{x}^{max}$. Thus,
\begin{eqnarray}
\hat{a}_{3}^{T}\vec{T}\hat{b}_{2}=(-\sqrt{\lambda_{x}^{max}},0,0)
\label{vector322a}
\end{eqnarray}
(iii) The vector $\hat{b}_{3}^{T}\vec{T}\hat{b}_{2}$ can be
simplified as
\begin{eqnarray}
\hat{b}_{3}^{T}\vec{T}\hat{b}_{2}&=&((\hat{b}_{3},T_{x}\hat{b}_{2}),(\hat{b}_{3},T_{y}\hat{b}_{2}),(\hat{b}_{3},T_{z}\hat{b}_{2}))
\nonumber\\&=&(-\|\hat{b}_{3}^{min}\|\|T_{x}\hat{b}_{2}\|,0,0)\nonumber\\&=&(-\sqrt{\lambda_{x}^{max}},0,0)
\label{vector333}
\end{eqnarray}
where $\hat{b}_{3}^{min}$ is the unit vector antiparallel to
$T_{x}\hat{b}_{2}$ and perpendicular to $T_{y}\hat{b}_{2}$ and
$T_{z}\hat{b}_{2}$. Since $\hat{b}_{3}^{min}$ is a unit vector,
 $\|\hat{b}_{3}^{min}\|=1$.\\
(iv) The vector $\hat{b}_{3}^{T}\vec{T}\hat{a}_{2}$ can be
simplified as
\begin{eqnarray}
\hat{b}_{3}^{T}\vec{T}\hat{a}_{2}&=&((\hat{b}_{3},T_{x}\hat{a}_{2}),(\hat{b}_{3},T_{y}\hat{a}_{2}),(\hat{b}_{3},T_{z}\hat{a}_{2}))
\nonumber\\&=&(0,\|\hat{b}_{3}^{max}\|\|T_{y}\hat{a}_{2}\|,0)=(0,\sqrt{\lambda_{y}^{max}},0)
\label{vector344}
\end{eqnarray}
where $\hat{b}_{3}^{max}$ is the unit vector along
$T_{y}\hat{a}_{2}$ and perpendicular to $T_{x}\hat{a}_{2}$ and
$T_{z}\hat{a}_{2}$.\\

From
(\ref{expectationvalue}),(\ref{vector311a}),(\ref{vector322a}),(\ref{vector333}),(\ref{vector344}),
we have
\begin{eqnarray}
\langle
B_{M}^{(8)}\rangle_{\rho}&=&max_{\hat{a}_{1},\hat{b}_{1}}[(\hat{a}_{1},(0,\sqrt{\lambda_{y}^{max}},0))
\nonumber\\&+&
(\hat{a}_{1},(\sqrt{\lambda_{x}^{max}},0,0))-(\hat{b}_{1},(0,\sqrt{\lambda_{y}^{max}},0))
\nonumber\\&+&(\hat{b}_{1},(\sqrt{\lambda_{x}^{max}},0,0))]\nonumber\\&=&\|\hat{a}_{1}^{max}\|\|(0,\sqrt{\lambda_{y}^{max}},0)\|
\nonumber\\&+&\|\hat{b}_{1}^{max}\|\|(0,\sqrt{\lambda_{y}^{max}},0)\|\nonumber\\&=&2\sqrt{\lambda_{y}^{max}}
\label{result31}
\end{eqnarray}
$\hat{a}_{1}^{max}$ is the unit vector parallel to
$(0,\sqrt{\lambda_{y}^{max}},0)$ and perpendicular to
$(\sqrt{\lambda_{x}^{max}},0,0)$; $\hat{b}_{1}^{max}$ is the unit
vectors antiparallel to $(0,\sqrt{\lambda_{y}^{max}},0)$ and
perpendicular to
$(\sqrt{\lambda_{x}^{max}},0,0)$.\\

\textbf{Case-III:} In this case we choose the vectors in such a way
that it maximizes the expectation value of the Mermin operator, given by
 $\langle B_{M}^{(9)}\rangle_{\rho}=2\sqrt{\lambda_{z}^{max}}$.\\
(i) The vector $\hat{a}_{3}^{T}\vec{T}\hat{a}_{2}$ can be
simplified as
\begin{eqnarray}
\hat{a}_{3}^{T}\vec{T}\hat{a}_{2}&=&((\hat{a}_{3},T_{x}\hat{a}_{2}),(\hat{a}_{3},T_{y}\hat{a}_{2}),(\hat{a}_{3},T_{z}\hat{a}_{2}))
\nonumber\\&=&(0,0,\|\hat{a}_{3}^{max}\|\|T_{z}\hat{a}_{2}\|)
\label{vector3111}
\end{eqnarray}
where $\hat{a}_{3}^{max}$ is the unit vector along
$T_{z}\hat{a}_{2}$ and perpendicular to $T_{x}\hat{a}_{2}$ and
$T_{y}\hat{a}_{2}$.
Since $\hat{a}_{3}^{max}$ is a unit vector,
$\|\hat{a}_{3}^{max}\|=1$. Again,
$\|T_{z}\hat{a}_{2}\|^{2}=(T_{z}\hat{a}_{2},T_{z}\hat{a}_{2})=(\hat{a}_{2},T_{z}^{T}T_{z}\hat{a}_{2})$.
If $\lambda_{z}^{max}$ is the largest eigenvalue of the symmetric
matrix $T_{z}^{T}T_{z}$ and $\hat{a}_{2}^{max}$ is the
corresponding unit vector then
$\|T_{z}\hat{a}_{2}\|^{2}=(\hat{a}_{2},T_{z}^{T}T_{z}\hat{a}_{2})=(\hat{a}_{2},\lambda_{z}^{max}\hat{a}_{2}^{max})$.
If $\hat{a}_{2}$ is the unit vector along $\hat{a}_{2}^{max}$, then
$\|T_{z}\hat{a}_{2}\|^{2}=\lambda_{z}^{max}$. Thus,
\begin{eqnarray}
\hat{a}_{3}^{T}\vec{T}\hat{a}_{2}=(0,0,\sqrt{\lambda_{z}^{max}})
\label{vector3111a}
\end{eqnarray}
(ii) The vector $\hat{a}_{3}^{T}\vec{T}\hat{b}_{2}$ can be
simplified as
\begin{eqnarray}
\hat{a}_{3}^{T}\vec{T}\hat{b}_{2}&=&((\hat{a}_{3},T_{x}\hat{b}_{2}),(\hat{a}_{3},T_{y}\hat{b}_{2}),(\hat{a}_{3},T_{z}\hat{b}_{2}))
\nonumber\\&=&(-\|\hat{a}_{3}^{min}\|\|T_{x}\hat{b}_{2}\|,0,0)
\label{vector3222}
\end{eqnarray}
where $\hat{a}_{3}^{min}$ is the unit vector antiparallel to
$T_{x}\hat{b}_{2}$ and perpendicular to $T_{y}\hat{b}_{2}$ and
$T_{z}\hat{b}_{2}$.
Since $\hat{a}_{3}^{min}$ is a unit vector,
$\|\hat{a}_{3}^{min}\|=1$. Again,
$\|T_{x}\hat{b}_{2}\|^{2}=(T_{x}\hat{b}_{2},T_{x}\hat{b}_{2})=(\hat{b}_{2},T_{x}^{T}T_{x}\hat{b}_{2})$.
If $\lambda_{x}^{max}$ is the largest eigenvalue of the symmetric
matrix $T_{x}^{T}T_{x}$ and $\hat{b}_{2}^{max}$ is the
corresponding unit vector then
$\|T_{x}\hat{b}_{2}\|^{2}=(\hat{b}_{2},T_{x}^{T}T_{x}\hat{b}_{2})=(\hat{b}_{2},\lambda_{x}^{max}\hat{b}_{2}^{max})$.
If $\hat{b}_{2}$ is the unit vector along $\hat{b}_{2}^{max}$ then
$\|T_{x}\hat{b}_{2}\|^{2}=\lambda_{x}^{max}$. Thus,
\begin{eqnarray}
\hat{a}_{3}^{T}\vec{T}\hat{b}_{2}=(-\sqrt{\lambda_{x}^{max}},0,0)
\label{vector3222a}
\end{eqnarray}
(iii) The vector $\hat{b}_{3}^{T}\vec{T}\hat{b}_{2}$ can be
simplified as
\begin{eqnarray}
\hat{b}_{3}^{T}\vec{T}\hat{b}_{2}&=&((\hat{b}_{3},T_{x}\hat{b}_{2}),(\hat{b}_{3},T_{y}\hat{b}_{2}),(\hat{b}_{3},T_{z}\hat{b}_{2}))
\nonumber\\&=&(-\|\hat{b}_{3}^{min}\|\|T_{x}\hat{b}_{2}\|,0,0)\nonumber\\&=&(-\sqrt{\lambda_{x}^{max}},0,0)
\label{vector3333}
\end{eqnarray}
where $\hat{b}_{3}^{min}$ is the unit vector antiparallel to
$T_{x}\hat{b}_{2}$ and perpendicular to $T_{y}\hat{b}_{2}$ and
$T_{z}\hat{b}_{2}$. Since $\hat{b}_{3}^{min}$ is a unit vector,
$\|\hat{b}_{3}^{min}\|=1$.\\
(iv) The vector $\hat{b}_{3}^{T}\vec{T}\hat{a}_{2}$ can be
simplified as
\begin{eqnarray}
\hat{b}_{3}^{T}\vec{T}\hat{a}_{2}&=&((\hat{b}_{3},T_{x}\hat{a}_{2}),(\hat{b}_{3},T_{y}\hat{a}_{2}),(\hat{b}_{3},T_{z}\hat{a}_{2}))
\nonumber\\&=&(0,0,\|\hat{b}_{3}^{max}\|\|T_{z}\hat{a}_{2}\|)=(0,0,\sqrt{\lambda_{z}^{max}})
\label{vector3444}
\end{eqnarray}
where $\hat{b}_{3}^{max}$ is the unit vector along
$T_{z}\hat{a}_{2}$ and perpendicular to $T_{x}\hat{a}_{2}$ and
$T_{y}\hat{a}_{2}$.\\

From
(\ref{expectationvalue}),(\ref{vector3111a}),(\ref{vector3222a}),(\ref{vector3333}),(\ref{vector3444}),
we have
\begin{eqnarray}
\langle
B_{M}^{(9)}\rangle_{\rho}&=&max_{\hat{a}_{1},\hat{b}_{1}}[(\hat{a}_{1},(0,0,\sqrt{\lambda_{z}^{max}}))
\nonumber\\&+&
(\hat{a}_{1},(\sqrt{\lambda_{x}^{max}},0,0))-(\hat{b}_{1},(0,0,\sqrt{\lambda_{z}^{max}}))
\nonumber\\&+&(\hat{b}_{1},(\sqrt{\lambda_{x}^{max}},0,0))]\nonumber\\&=&\|\hat{a}_{1}^{max}\|\|(0,0,\sqrt{\lambda_{z}^{max}})\|
\nonumber\\&+&\|\hat{b}_{1}^{max}\|\|(0,0,\sqrt{\lambda_{z}^{max}})\|\nonumber\\&=&2\sqrt{\lambda_{z}^{max}}
\label{result31}
\end{eqnarray}
$\hat{a}_{1}^{max}$ is the unit vector parallel to
$(0,0,\sqrt{\lambda_{z}^{max}})$ and perpendicular to
$(\sqrt{\lambda_{x}^{max}},0,0)$; $\hat{b}_{1}^{max}$ is the unit
vectors antiparallel to $(0,0,\sqrt{\lambda_{z}^{max}})$ and
perpendicular to
$(\sqrt{\lambda_{x}^{max}},0,0)$.\\
Thus, the maximum expectation value of the Mermin operator with
respect to the state $\rho$ is given by
\begin{eqnarray}
\langle B_{M}^{max}\rangle_{\rho}&=&max\{\langle
B_{M}^{(7)}\rangle_{\rho},\langle
B_{M}^{(8)}\rangle_{\rho},\langle
B_{M}^{(9)}\rangle_{\rho}\}\nonumber\\
&=&max\{4\sqrt{\lambda_{x}^{max}},2\sqrt{\lambda_{y}^{max}}
,2\sqrt{\lambda_{z}^{max}}\}
\end{eqnarray}
Mermin inequality is violated if
\begin{eqnarray}
\langle
B_{M}^{max}\rangle_{\rho}=max\{4\sqrt{\lambda_{x}^{max}},2\sqrt{\lambda_{y}^{max}},2\sqrt{\lambda_{z}^{max}}\}>2
\end{eqnarray}
Thus
$max\{2\sqrt{\lambda_{x}^{max}},\sqrt{\lambda_{y}^{max}},\sqrt{\lambda_{z}^{max}}\}>1$.
Hence, proved.\\

\textbf{Corollary-1:} If $T_{y}^{T}T_{y}$ has two equal largest
eigenvalues $\lambda_{y}^{max}$ and $T_{x}^{T}T_{x}$ and
$T_{z}^{T}T_{z}$ have unique largest eigenvalues $\lambda_{x}^{max}$
and $\lambda_{z}^{max}$ respectively, then Mermin's inequality is
violated if
\begin{eqnarray}
\langle
B_{M}^{max}\rangle_{\rho}&=&max\{2\sqrt{\lambda_{x}^{max}}},4\sqrt{\lambda_{y}^{max}},
2\sqrt{\lambda_{z}^{max}\}\nonumber\\&>&2\label{corollary1}
\end{eqnarray}

\textbf{Corollary-2:} If $T_{z}^{T}T_{z}$ has two equal largest
eigenvalues $\lambda_{z}^{max}$, and $T_{x}^{T}T_{x}$ and
$T_{y}^{T}T_{y}$ have unique largest eigenvalue $\lambda_{x}^{max}$
and $\lambda_{y}^{max}$ respectively, then Mermin's inequality is
violated if
\begin{eqnarray}
\langle
B_{M}^{max}\rangle_{\rho}&=&max\{2\sqrt{\lambda_{x}^{max}},2\sqrt{\lambda_{y}^{max}},
4\sqrt{\lambda_{z}^{max}}\}\nonumber\\&>&2\label{corollary2}
\end{eqnarray}

\textbf{Corollary-3:} If $T_{x}^{T}T_{x}$ and $T_{y}^{T}T_{y}$ have
two equal largest eigenvalues $\lambda_{x}^{max}$ and
$\lambda_{y}^{max}$  respectively, and $T_{z}^{T}T_{z}$ has unique largest
eigenvalue $\lambda_{z}^{max}$,  then Mermin's
inequality is violated if
\begin{eqnarray}
\langle
B_{M}^{max}\rangle_{\rho}&=&max\{4\sqrt{\lambda_{x}^{max}},4\sqrt{\lambda_{y}^{max}},
2\sqrt{\lambda_{z}^{max}}\}\nonumber\\&>&2\label{corollary3}
\end{eqnarray}

\textbf{Corollary-4:} If $T_{x}^{T}T_{x}$ and $T_{z}^{T}T_{z}$ have
two equal largest eigenvalue $\lambda_{x}^{max}$ and
$\lambda_{z}^{max}$,  respectively and $T_{y}^{T}T_{y}$ has unique largest
eigenvalue $\lambda_{y}^{max}$, then Mermin's
inequality is violated if
\begin{eqnarray}
\langle
B_{M}^{max}\rangle_{\rho}&=&max\{4\sqrt{\lambda_{x}^{max}},2\sqrt{\lambda_{y}^{max}},
4\sqrt{\lambda_{z}^{max}}\}\nonumber\\&>&2\label{corollary4}
\end{eqnarray}

\textbf{Corollary-5:} If $T_{y}^{T}T_{y}$ and $T_{z}^{T}T_{z}$ have
two equal largest eigenvalue $\lambda_{y}^{max}$ and
$\lambda_{z}^{max}$  respectively, and $T_{x}^{T}T_{x}$ has unique largest
eigenvalue $\lambda_{x}^{max}$, then Mermin's
inequality is violated if
\begin{eqnarray}
\langle
B_{M}^{max}\rangle_{\rho}&=&max\{2\sqrt{\lambda_{x}^{max}},4\sqrt{\lambda_{y}^{max}},
4\sqrt{\lambda_{z}^{max}}\}\nonumber\\&>&2\label{corollary5}
\end{eqnarray}

\section{Examples}

\textbf{Example-1:} The generalized GHZ state can be written in
terms of Pauli matrices as
\begin{eqnarray}
\rho_{GGHZ}&=&\frac{1}{8}[I\otimes I\otimes I+I\otimes
\sigma_{z}\otimes \sigma_{z}+\sigma_{z}\otimes I\otimes
\sigma_{z}\nonumber\\&+&\sigma_{z}\otimes \sigma_{z}\otimes I
+(\alpha^{2}-\beta^{2})(\sigma_{z}\otimes I\otimes
I\nonumber\\&+&I\otimes \sigma_{z}\otimes I+I\otimes I\otimes
\sigma_{z} )+2\alpha\beta(\sigma_{x}\otimes \sigma_{x}\otimes
\sigma_{x}\nonumber\\&-&\sigma_{x}\otimes \sigma_{y}\otimes
\sigma_{y}-\sigma_{y}\otimes \sigma_{x}\otimes
\sigma_{y}\nonumber\\&-&\sigma_{y}\otimes \sigma_{y}\otimes
\sigma_{x})],~~\alpha^{2}+\beta^{2}=1\label{gghz}
\end{eqnarray}
The matrices $T_{x}^{T}T_{x}$,$T_{y}^{T}T_{y}$ are given by
\begin{eqnarray}
&&T_{x}^{T}T_{x}=T_{y}^{T}T_{y}=\begin{pmatrix}
  4\alpha^{2}\beta^{2} & 0 & 0\\
  0 & 4\alpha^{2}\beta^{2} & 0 \\
  0 & 0 & 0
\end{pmatrix}
\label{symmat1}
\end{eqnarray}
The matrix $T_{z}^{T}T_{z}$ is a zero matrix.
The largest eigenvalues of the matrices
$T_{x}^{T}T_{x}$,$T_{y}^{T}T_{y}$ are given by
\begin{eqnarray}
&&\lambda_{x}^{max}=4\alpha^{2}\beta^{2}\nonumber\\&&
\lambda_{y}^{max}=4\alpha^{2}\beta^{2}\nonumber\\&&
\label{lareig1}
\end{eqnarray}
The maximum expectation value of the Mermin operator with respect
to the state $\rho_{GGHZ}$ is given by
\begin{eqnarray}
\langle B_{M}^{max}\rangle_{\rho_{GGHZ}}=8\alpha\beta \label{max1}
\end{eqnarray}
Therefore, the generalized GHZ state violates Mermin's inequality if
\begin{eqnarray}
 2\alpha\beta >\frac{1}{2}\label{vio1}
\end{eqnarray}
The same result has been found numerically by Scarani and Gisin
\cite{scarani}.

\textbf{Example-2:} Let us consider a pure state which is a
coherent superposition of the $W-$ state and a separable state
$|000\rangle$. The superposed state can be expressed as
\begin{eqnarray}
|\Psi\rangle_{W,S}=\sqrt{1-p}|W\rangle+\sqrt{p}|000\rangle,~~0\leq
p\leq 1 \label{super1}
\end{eqnarray}
where
$|W\rangle=\frac{1}{\sqrt{3}}(|001\rangle+|010\rangle+|100\rangle)$.
In this case the symmetric matrices
$T_{x}^{T}T_{x}$,$T_{y}^{T}T_{y}$,$T_{z}^{T}T_{z}$ take the form
\begin{eqnarray}
&&T_{x}^{T}T_{x}=\begin{pmatrix}
  \frac{4}{9}(1-p)^{2} & 0 & \frac{4}{3\sqrt{3}}\sqrt{p}(1-p)^{\frac{3}{2}} \\
  0 & 0 & 0 \\
  \frac{4}{3\sqrt{3}}\sqrt{p}(1-p)^{\frac{3}{2}} & 0 &
  \frac{4}{9}(1-p)(1+2p)
\end{pmatrix} \nonumber\\&&
T_{y}^{T}T_{y}=\begin{pmatrix}
  0 & 0 & 0 \\
  0 & \frac{4}{9}(1-p)^{2} & 0 \\
  0 & 0 & \frac{4}{9}(1-p)^{2}
\end{pmatrix}\nonumber\\&&
T_{z}^{T}T_{z}=\begin{pmatrix}
  \frac{4}{9}(1-p)(1+2p) & 0 & f \\
  0 & \frac{4}{9}(1-p)^{2} & 0 \\
  f& 0 & g
\end{pmatrix}
\label{symmat2}
\end{eqnarray}
where
$f=\frac{4}{3\sqrt{3}}\sqrt{p}(1-p)^{\frac{3}{2}}+\frac{2}{\sqrt{3}}\sqrt{p}\sqrt{1-p}(2p-1)$
and $g=\frac{4}{3}p(1-p)+(2p-1)^{2}$.
The largest eigenvalues of the matrices
$T_{x}^{T}T_{x}$,$T_{y}^{T}T_{y}$,$T_{z}^{T}T_{z}$ are given by
\begin{eqnarray}
&&\lambda_{x}^{max}=(1-p)(\frac{4}{9}+\frac{2}{9}p+\frac{2}{9}\sqrt{12p-3p^{2}})\nonumber\\&&
\lambda_{y}^{max}=\frac{4}{9}(1-p)^{2}\nonumber\\&&
\lambda_{z}^{max}=\frac{1}{18}\sqrt{256p^{4}-640p^{3}+672p^{2}-232p+25}\nonumber\\&&+\frac{13}{18}+\frac{8}{9}p^{2}-\frac{10}{9}p
\label{lareig2}
\end{eqnarray}
The maximum expectation value of the Mermin operator with respect
to the state $|\Psi\rangle_{W,S}$ is given by
\begin{eqnarray}
&&\langle
B_{M}^{max}\rangle_{|\Psi\rangle_{W,S}\langle\Psi|}=4\sqrt{\lambda_{y}^{max}}~~0\leq
p\leq0.43\nonumber\\&&\langle
B_{M}^{max}\rangle_{|\Psi\rangle_{W,S}\langle\Psi|}=2\sqrt{\lambda_{x}^{max}},~~0.43\leq
p\leq0.45\nonumber\\&&\langle
B_{M}^{max}\rangle_{|\Psi\rangle_{W,S}\langle\Psi|}=2\sqrt{\lambda_{z}^{max}},~~0.45\leq
p\leq1 \label{max2}
\end{eqnarray}
It can be easily seen from figs.1(a), fig.1(b), fig.1(c) that
the state $|\Psi\rangle_{W,S}$
violates Mermin's inequality when $0\leq p\leq 0.25$. This result was
obtained in \cite{chi}.

\begin{figure}[!ht]
\resizebox{7cm}{4cm}{\includegraphics{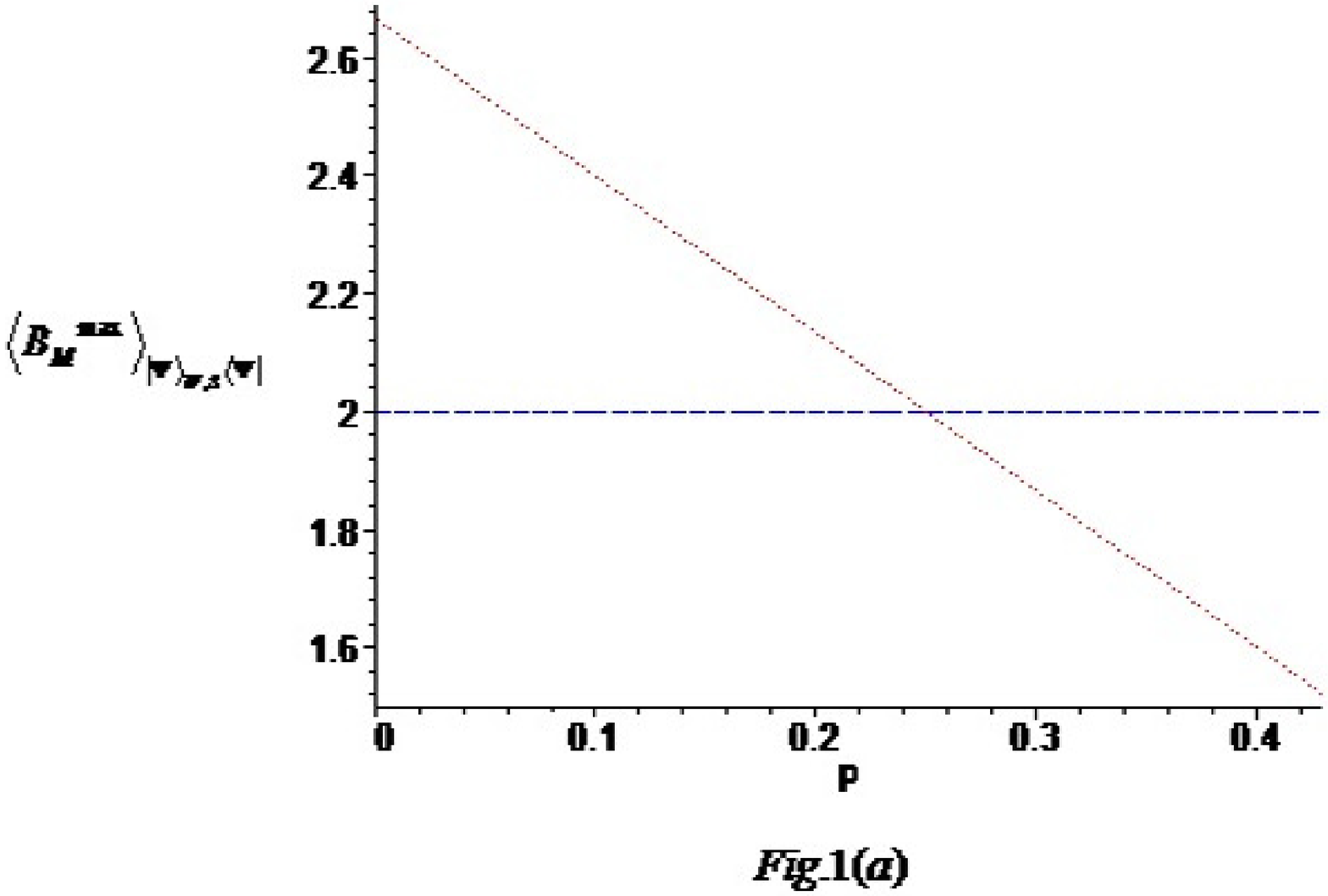}}
\end{figure}
\begin{figure}[!ht]
\resizebox{7cm}{4cm}{\includegraphics{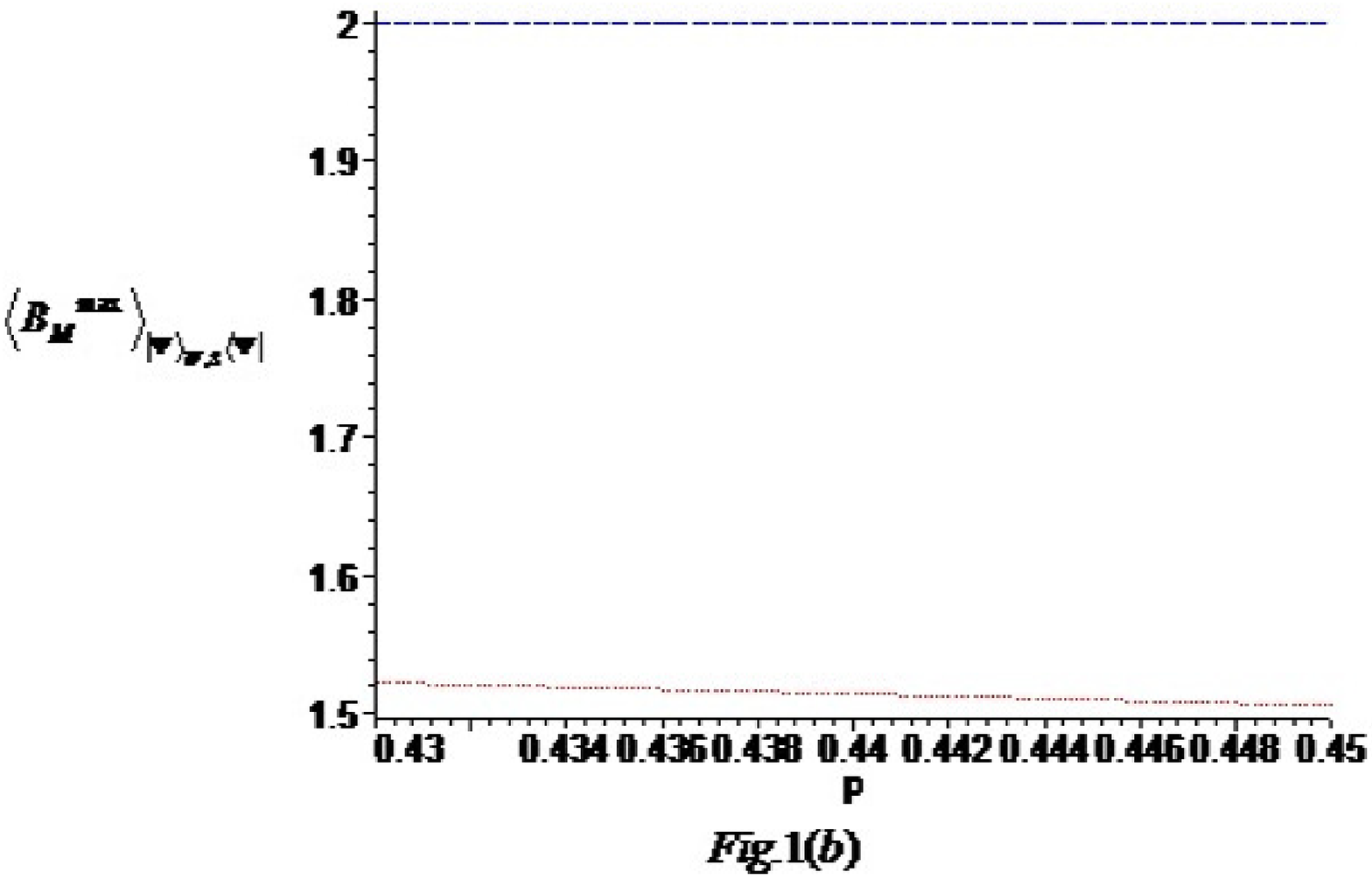}}
\end{figure}
\begin{figure}[!ht]
\resizebox{7cm}{4cm}{\includegraphics{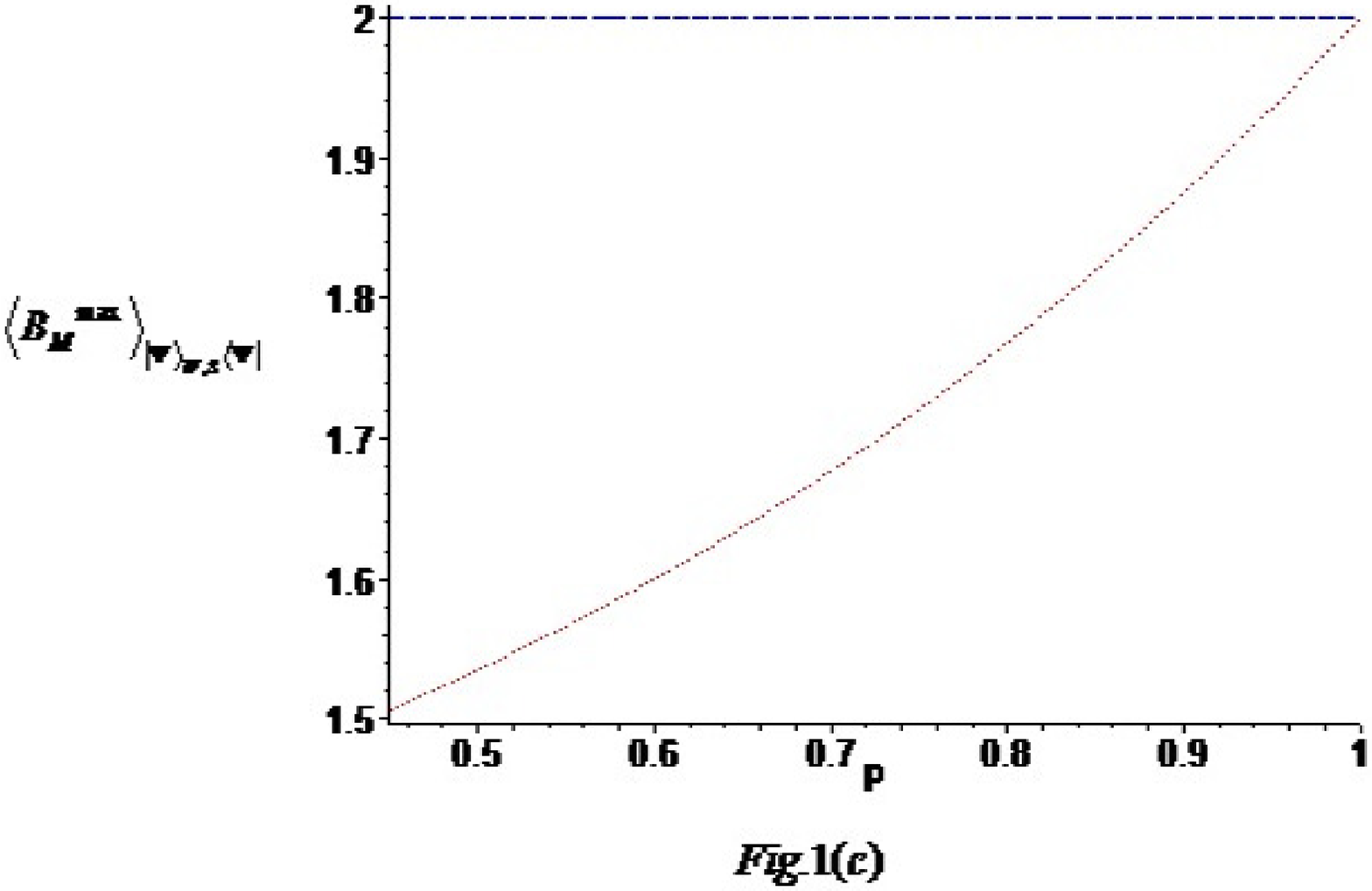}}
\caption{\footnotesize Violation of the Mermin inequality versus
the state parameter $p$ for the pure state given by
Eq.(\ref{super1})  }
\end{figure}



\textbf{Example-3:} Let us consider a mixed state $\varrho$ which
is described by the density operator
\begin{eqnarray}
\varrho=p|\psi\rangle_{GHZ}\langle\psi|+(1-p)|\psi\rangle_{W}\langle\psi|
\label{mixedstate}
\end{eqnarray}
where
$|\psi\rangle_{GHZ}=\frac{1}{\sqrt{2}}(|000\rangle+|111\rangle)$
and
$|\psi\rangle_{W}=\frac{1}{\sqrt{3}}(|001\rangle+|010\rangle+|100\rangle)$.
Our task is to find out the range of the parameter $p$ for which
$\varrho$ violates the Mermin inequality.  Hence, we have
to calculate the largest eigenvalues of the symmetric matrices
$T_{x}^{T}T_{x}$,$T_{y}^{T}T_{y}$,$T_{z}^{T}T_{z}$. Since in this
case the matrices $T_{y}^{T}T_{y}$,$T_{z}^{T}T_{z}$ have two equal
largest eigenvalues, and $T_{x}^{T}T_{x}$ has a unique largest
eigenvalue, one has
\begin{eqnarray}
\langle
B_{M}^{max}\rangle_{\varrho}&=&max\{2\sqrt{\lambda_{x}^{max}},4\sqrt{\lambda_{y}^{max}},4\sqrt{\lambda_{z}^{max}}\}\nonumber\\\label{examplemax}
\end{eqnarray}
where
$\lambda_{x}^{max}=\frac{4}{9}-\frac{8}{9}p+\frac{17}{18}p^{2}+\frac{1}{6}\sqrt{25p^{4}-32p^{3}+16p^{2}}$,
$\lambda_{y}^{max}=\frac{4}{9}-\frac{8}{9}p+\frac{13}{9}p^{2}$,
and $\lambda_{z}^{max}=(1-p)^{2}$.\\
It follows from Eq.(\ref{examplemax}) that
\begin{eqnarray}
\langle B_{M}^{max}\rangle_{\varrho}&=&4(1-p),~~0\leq p\leq 0.43
\nonumber\\&=& 4\sqrt{\frac{4}{9}-\frac{8}{9}p+\frac{13}{9}p^{2}},
0.43\leq p\leq 1 \label{examplemax1}
\end{eqnarray}
Fig.2(a) and Fig.2(b) clearly show that the mixed state $\varrho$
violates Mermin's inequality for all values of the parameter $p$, i.e., $0\leq p\leq 1$.

\begin{figure}[!ht]
\resizebox{7cm}{4cm}{\includegraphics{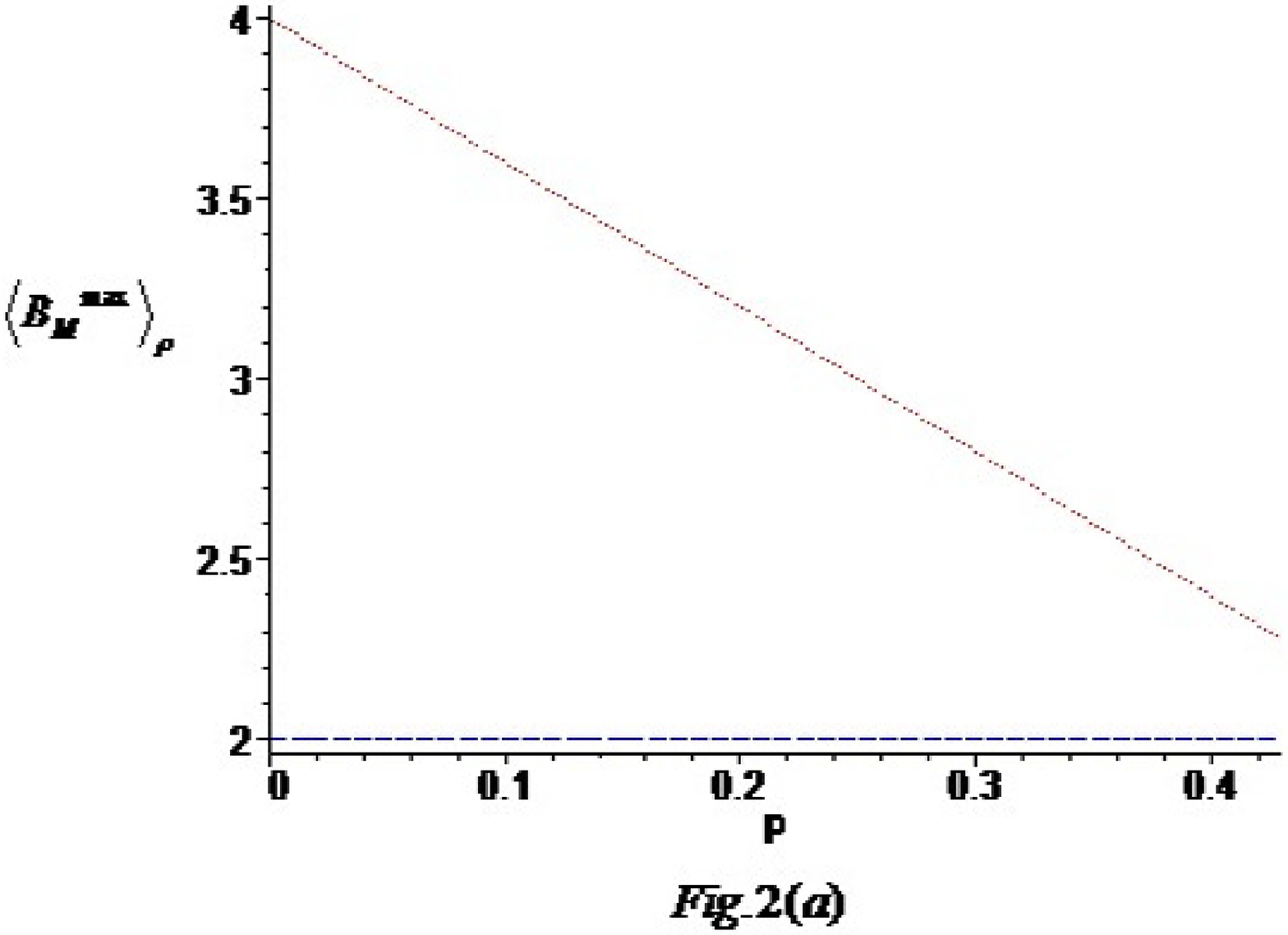}}
\end{figure}
\begin{figure}[!ht]
\resizebox{7cm}{4cm}{\includegraphics{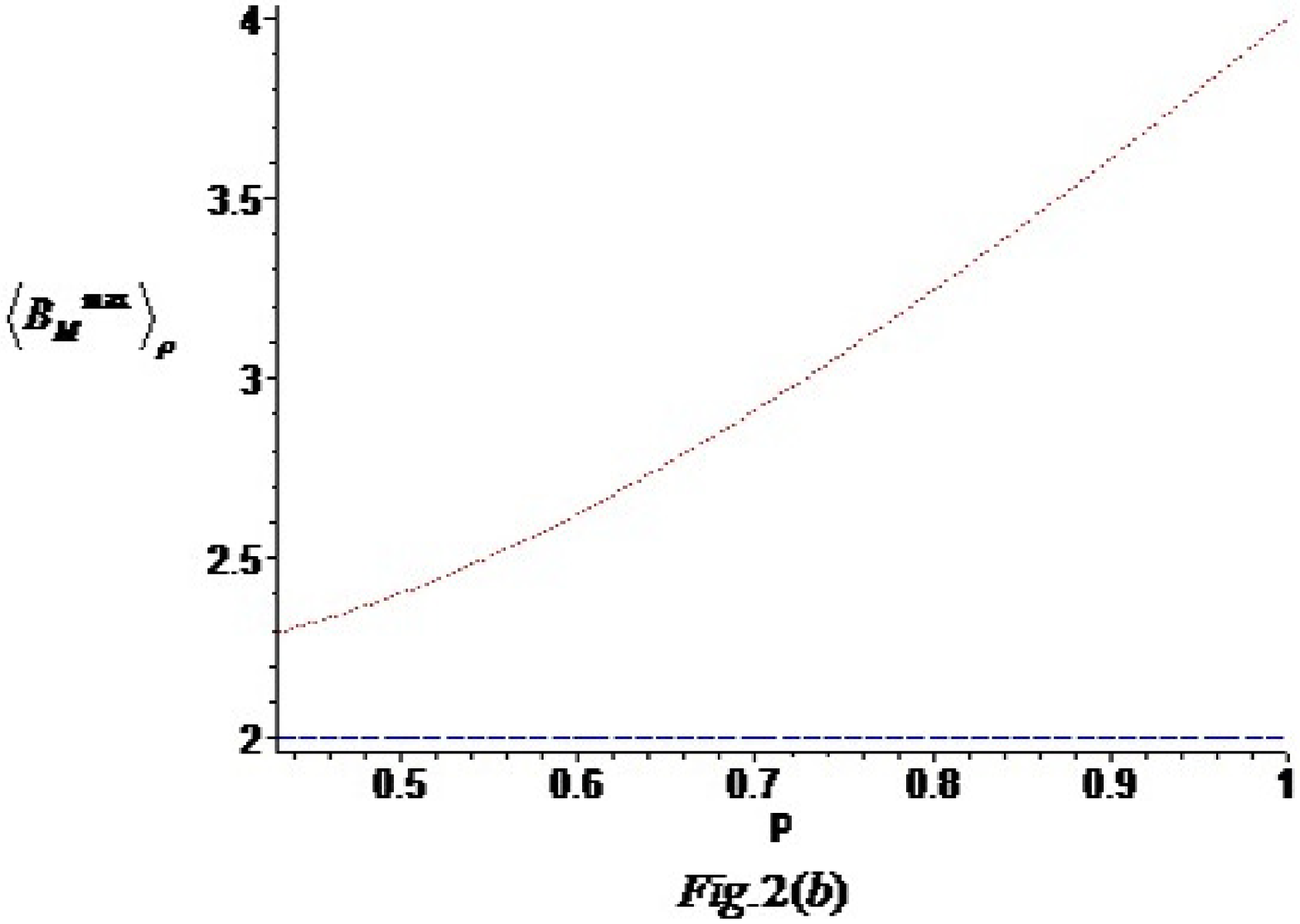}}
\caption{\footnotesize Violation of the Mermin inequality versus
the state parameter $p$ for the mixed state given by
Eq.(\ref{mixedstate})}
\end{figure}



\section{Conclusions}

In this work we have studied  Mermin's inequality for three qubit
states, the violation of which predicts the existence of quantum
correlations between the outcomes of the distant measurements on
three qubit systems. Prior to this work there did not exist in the literature
any closed form expression that gives the maximal violation of the
Mermin inequality. Motivated by the analogous criterion for two qubit systems
\cite{horo},  we have here presented some analytical formulae in
terms of eigenvalues of symmetric matrices, that provide conditions for violating the
Mermin inequality by both pure and mixed arbitrary three qubit states.
Our results are useful
in obtaining the violation of Mermin's inequality because using them one does not need to
perform optimization procedures over spin  measurements in all possible directions.
We have illustrated our results with a few examples of pure and mixed states, confirming
the range of violation of the Mermin inequality obtained in
earlier works \cite{scarani1, chi}.

{\it Acknowledgements}: ASM acknowledges support from the project SR/S2/LOP-08/2013
of DST, India.

\end{document}